\documentclass[a4paper,fleqn,usenatbib]{mnras}

\usepackage{newtxtext,newtxmath}
\usepackage[T1]{fontenc}
\usepackage{ae,aecompl}

\usepackage{graphicx}
\usepackage{amsmath}
\usepackage{amssymb}
\usepackage{bm}
\usepackage{comment}

\newcommand{\beq}{\begin{equation}}
\newcommand{\eeq}{\end{equation}}
\newcommand{\beqa}{\begin{eqnarray}}
\newcommand{\eeqa}{\end{eqnarray}}
\newcommand{\simgt}{\lower.5ex\hbox{$\; \buildrel > \over \sim \;$}}
\newcommand{\simlt}{\lower.5ex\hbox{$\; \buildrel < \over \sim \;$}}

\newcommand{\vir}{\mathrm{vir}}
\newcommand{\Msun}{\mathrm{M}_\odot}

\title[Cross-correlation of the tSZ and WL]
{Investigating cluster astrophysics and cosmology
with cross-correlation of the thermal Sunyaev--Zel'dovich effect and weak lensing}

\author[K. Osato et al.]{
Ken Osato,$^{1,8}$\thanks{E-mail: ken.osato@utap.phys.s.u-tokyo.ac.jp (KO)}
Samuel Flender,$^{2,3}$
Daisuke Nagai,$^{4,5}$
Masato Shirasaki,$^{6,8}$
\newauthor
and Naoki Yoshida$^{1,7,8}$
\\
$^{1}$Department of Physics, School of Science, The University of Tokyo,
7-3-1 Hongo, Bunkyo, Tokyo, 113-0033, Japan\\
$^{2}$HEP Division, Argonne National Laboratory, 9700 S. Cass Ave., Lemont, IL 60439, USA\\
$^{3}$Kavli Institute for Cosmological Physics, The University of Chicago, Chicago, IL 60637, USA\\
$^{4}$Department of Physics, Yale University, New Haven, CT 06520, USA\\
$^{5}$Yale Center for Astronomy and Astrophysics, Yale University, New Haven, CT 06520,USA\\
$^{6}$National Astronomical Observatory of Japan, Mitaka, Tokyo, 181-8588, Japan\\
$^{7}$Kavli Institute for the Physics and Mathematics of the Universe (WPI),
U-Tokyo Institutes for Advanced Study,\\
The University of Tokyo, Kashiwa, Chiba, 277-8583, Japan\\
$^{8}$CREST, Japan Science and Technology Agency,
4-1-8 Honcho, Kawaguchi, Saitama, 332-0012, Japan
}

\date{Accepted XXX. Received YYY; in original form ZZZ}

\pubyear{2017}

\begin{document}
\label{firstpage}
\pagerange{\pageref{firstpage}--\pageref{lastpage}}
\maketitle

\begin{abstract}
Recent detections of the cross-correlation of
the thermal Sunyaev--Zel'dovich (tSZ) effect and
weak gravitational lensing (WL) enable
unique studies of cluster astrophysics and cosmology.
In this work, we present constraints on the amplitude of
the non-thermal pressure fraction in galaxy clusters, $\alpha_0$,
and the amplitude of the matter power spectrum, $\sigma_8$,
using measurements of the tSZ power spectrum from {\it Planck},
and the tSZ-WL cross-correlation from {\it Planck}
and the Red Cluster Sequence Lensing Survey.
We fit the data to a semi-analytic model with the covariance matrix
using $N$-body simulations.
We find that the tSZ power spectrum alone prefers $\sigma_8 \sim 0.85$ and
a large fraction of non-thermal pressure ($\alpha_0 \sim 0.2$--$0.3$).
The tSZ-WL cross-correlation on the other hand prefers a significantly
lower $\sigma_8 \sim 0.6$, and low $\alpha_0 \sim 0.05$.
We show that this tension can be mitigated by allowing
for a steep slope in the stellar-mass-halo-mass relation,
which would cause a reduction of the gas in low-mass halos.
In such a model, the combined data prefer $\sigma_8 \sim 0.7$ and $\alpha_0 \sim 0.2$,
consistent with predictions from hydrodynamical simulations.
\end{abstract}




\section{Introduction}
\label{sec:introduction}
The observation of the anisotropy of cosmic microwave background (CMB)
provides us with generous information of our Universe.
The Sunyaev--Zel'dovich (SZ) effect \citep{Sunyaev1972, Sunyaev1980}
is one of the effects which give rise to
the anisotropy after CMB photons decouple with the hot plasma.
As CMB photons are scattered by hot electrons,
energy transfer occurs via Compton scattering.
As a result, the energy spectrum of CMB deviates from the black-body spectrum.
There are two types of SZ effects, one is thermal SZ effect (tSZ),
which is due to the thermal motion of hot gas, and the other one is
kinetic SZ effect (kSZ), which is due to the bulk motion of gas.
Since the hot electrons sourcing the tSZ signal originate predominantly from massive halos,
the tSZ signal reflects thermodynamic properties of intracluster medium (ICM).
Though SZ effects have been important probes into the structure formation
in the Universe and astrophysics of the ICM,
the measurement of SZ effects is challenging because of the small amplitude
of the signal and foreground contamination.
Due to significant improvement in resolution and sensitivity,
several CMB experiments have been able to detect the tSZ effect
\citep[see, e.g., ][]{Hasselfield2013, Bleem2015, Planck2015tSZ}.
In order to make full use of the recent observations,
the accurate and precise modeling of SZ effects is essential for cosmology.

There are various methods for modeling the tSZ effect.
One of the methods is the analytical modeling
of radial profiles of gas density and pressure
\citep{Komatsu2001, Komatsu2002, Ostriker2005}.
Then, one can obtain the tSZ power spectrum using the halo model
\citep{Cole1988, Komatsu1999}.
However, the evolution of gas is governed by complex baryonic physics,
e.g., star formation, feedback, and radiative cooling,
which are difficult to model analytically.
This difficulty directly leads to the uncertainty of the model.
One of the solutions to take baryonic physics into account properly
is employing hydrodynamical simulations.
The gas pressure profile of halos with different masses and redshifts
can be measured from cosmological hydrodynamical simulations
which include baryonic physics and the obtained profile is applied
to model the tSZ signal based on a halo model
\citep{Battaglia2010, Battaglia2012}.
Alternatively, one can also obtain the gas pressure field in the Universe from
hydrodynamical simulation, and hence the tSZ signal directly by integrating the
pressure field in the line-of-sight direction
\citep[see, e.g.,][]{McCarthy2014, Dolag2016}.

However, running hydrodynamical simulations is computationally expensive, and
the volume covered by the simulation is limited.
To overcome these problems, a realistic solution is the {\it semi-analytic}
prescription which combines analytical modeling and $N$-body simulations.
In \citet{Sehgal2010, Trac2011}, the authors run $N$-body simulations
and create halo catalogs from the simulations.
From the halo mass, the gas pressure profile of the halo is obtained analytically,
and then gas pressure is {\it pasted} onto each particle
in the $N$-body simulation.
Dark matter only simulations are computationally more efficient,
and can therefore be used to cover larger volumes than hydrodynamical simulations.

Furthermore, in this method we can incorporate various factors
which are not taken into account in the analytical models,
e.g., asphericity of halos or the effects of substructures.
Another way to model the pressure profile
is to make use of X-ray or tSZ observations
\citep{Arnaud2010, PlanckCollaboration2013}.
Since the pressure profile can be inferred from these observations,
we can learn about the relation between pressure and cluster mass and redshift.
Though such observations are typically limited at the low redshift,
substantial fraction of the tSZ power spectrum comes from the high redshift
clusters. This fact leads to the uncertainty in the modeling.

In addition to tSZ, we focus on weak gravitational lensing (WL)
by the large-scale structure, so-called comic shear
\citep[for a review, see][]{Bartelmann2001, Munshi2008, Kilbinger2015}.
The path of photons from distant galaxies is distorted
by gravitational potential of intervening matter.
WL reflects the abundance of matter in the line-of-sight direction
and thus can be a promising probe into the nature of dark matter and
dark energy.
Unlike the tSZ, WL is mostly determined by gravity and
less affected by baryonic physics. The nonlinear evolution by gravity
is well modeled by $N$-body simulations.

The information that can be extracted either from WL or tSZ is limited;
WL suffers from degeneracies between cosmological parameters,
while tSZ suffers from astrophysical uncertainties.
Thus, a combination of WL and tSZ can arguably be
more efficient in extracting cosmological parameters.
For this purpose, we focus on cross-correlation of WL and tSZ
in addition to the auto-power spectrum of the tSZ.
The cross-correlation analysis has a possibility to
enable us to place more stringent constraints on cosmological parameters
evading the astrophysical uncertainties and implications to physics of ICM
\citep{Munshi2014, Ma2015, Hojjati2015, Battaglia2015}.
Furthermore, the cross-correlation has already been detected by several groups
\citep[][note that their Compton-$y$ maps are based on {\it Planck} data
but constructed in different ways.]{VanWaerbeke2014, Hill2014, Hojjati2017}
and is one of scientific goals of current and forthcoming surveys, e.g.\ the
Hyper Suprime-Cam survey \citep[HSC;][]{Aihara2017}
\footnote{\url{http://hsc.mtk.nao.ac.jp/ssp/}},
Dark Energy Survey \citep[DES;][]{DarkEnergySurveyCollaboration2016}
\footnote{\url{http://www.darkenergysurvey.org/}},
and Large Synoptic Survey Telescope \citep[LSST;][]{LSST2009}
\footnote{\url{http://www.lsst.org/}} for WL, and
Atacama Cosmology Telescope \citep[ACT/ACTPol;][]{Niemack2010, Swetz2011}
and South Pole Telescope \citep[SPT/SPTPol;][]{Carlstrom2011, Austermann2012}
for tSZ.

It is timely to investigate the ability of the cross-correlation
with numerical simulations.
In this paper, we combine the output from an $N$-body simulation
with a semi-analytic model for the pressure,
in order to create mock tSZ and WL maps.
Using these maps, we estimate the covariance matrix of the tSZ power spectrum
and the tSZ-WL cross-correlation.
The main results of this analysis are the constrains on $\sigma_8$ and
the amount of non-thermal pressure in the ICM, derived from recent measurements
from {\it Planck} \citep{Planck2015tSZ}
and the Red Cluster Sequence Lensing Survey (RCSLenS)
\citep{Hildebrandt2016, Hojjati2017}.
Recently, it is reported that there is a tension of inferred $\sigma_8$
between the measurements of CMB temperature anisotropy and
large scale structure, e.g., power spectrum of WL \citep{Battye2015, Leauthaud2017}.
Furthermore, tSZ power spectra measured from SPT and ACT are lower than
the prediction based on {\it Planck} best-fit cosmological parameters \citep{Planck2015tSZ}.
This fact also may be related to the $\sigma_8$ tension,
though there is a possibility that the incomplete separation of foreground contamination
causes the low amplitude of the power spectrum.
The tSZ-WL cross-correlation along with tSZ power spectrum is
one of promissing probes into this problem.

This paper is organized as follows.
In Section~\ref{sec:formalism}, we review the basics of tSZ and WL,
and the analytic halo model for power spectra and cross-correlation.
We describe our semi-analytic model and simulations
in Section~\ref{sec:methods}.
In Section~\ref{sec:results}, we present measured spectra and
cross-correlation obtained from our model and
constraints on the property of non-thermal pressure and $\sigma_8$.
We conclude in Section~\ref{sec:conclusions}.

Throughout this paper, we assume the Universe is spatially flat
and follows the $\Lambda$CDM model.
We adopt cosmological parameters inferred from
temperature and polarization data set of CMB (TT,TE,EE+lowP)
from the {\it Planck} mission \citep{Planck2015parameters}.
The relative energy density of matter, baryons, and cosmological constant
at the present Universe are $\Omega_\mathrm{m} = 0.3156$,
$\Omega_\mathrm{b} = 0.04917$, $\Omega_\Lambda = 0.6844$.
The Hubble parameter is $h=0.6727$ with $H_0=100h\,\mathrm{km/s/Mpc}$.
The slope and amplitude of the scalar pertubation are
$n_\mathrm{s} = 0.9645$ and $A_\mathrm{s} = 2.2065 \times 10^{-9}$
with the pivot scale $k_\mathrm{pivot} = 0.05\,\mathrm{Mpc}^{-1}$.
Though we will constrain the amplitude later in this paper,
the fiducial value of the amplitude at the scale of $8\,\mathrm{Mpc}/h$
is $\sigma_8 = 0.831$.

\section{Formalism}
\label{sec:formalism}

\subsection{The thermal Sunyaev--Zel'dovich effect}
\label{sec:tSZ}
Here, we briefly review basic equations of tSZ effect
in the non-relativistic regime
\citep[for detailed reviews, see e.g.,][]{Birkinshaw1999, Carlstrom2002, Kitayama2014}.
The variation of temperature scales as the
line-of-sight integration of the electron pressure $P_\mathrm{e}$,
\beq
\label{eq:y_def}
\frac{\Delta T}{T_\mathrm{CMB}} = g_\nu (x) y = g_\nu (x) \frac{\sigma_\mathrm{T}}{m_\mathrm{e}c^2}
\int P_\mathrm{e} \, dl ,
\eeq
where $y$ is Compton-$y$ parameter, $T_\mathrm{CMB} = 2.726\,\mathrm{K}$
is the CMB temperature,
$\sigma_\mathrm{T}$ is the Thomson scattering cross-section,
$m_\mathrm{e}$ is the electron mass, and $g_\nu(x)$ is the frequency
dependent function given by
\beq
g_\nu (x) = x\frac{e^x-1}{e^x+1}-4, \, x = \frac{h\nu}{k_\mathrm{B}T} .
\eeq
We do not include relativistic corrections for $g_\nu (x)$
\citep{Itoh1998, Nozawa1998} because this effect is subdominant
in our interested scales and we basically focus on Compton-$y$.
For fully ionized primordial gas, the electron pressure $P_\mathrm{e}$
is related with thermal pressure $P_\mathrm{th}$ as
\beq
P_\mathrm{e} = \frac{2X+2}{5X+3} P_\mathrm{th},
\eeq
where $X=0.76$ is the hydrogen mass fraction.
The main task is to construct a thermal gas pressure profile model
from an analytical prescription, or observation.

The observable of tSZ is Compton-$y$ parameter and
the power spectrum of Compton-$y$ is the fundamental statistic for tSZ.
Here, let us consider the basic scheme of computing the power spectrum
based on the halo model.
Following the halo model formalism in \citet{Cole1988, Komatsu1999},
we can derive the expression for the angular power spectrum
of Compton-$y$ as the sum of 1-halo and 2-halo contributions,
\beqa
\label{eq:yy_hm1}
C^{yy} (\ell) &=& C^{yy(\mathrm{1h})} (\ell) + C^{yy(\mathrm{2h})} (\ell) , \\
C^{yy(\mathrm{1h})} (\ell) &=& \int_0^{z_\mathrm{dec}} dz\,
\frac{d^2 V}{dzd\Omega} \nonumber \\
\label{eq:yy_hm2}
&& \times \int_{M_\mathrm{min}}^{M_\mathrm{max}} dM\, \frac{dn(M, z)}{dM}
|y_\ell (M, z)|^2 , \\
C^{yy(\mathrm{2h})} (\ell) &=& \int_0^{z_\mathrm{dec}} dz\, \frac{d^2 V}{dzd\Omega}
P_\mathrm{m} (k_\ell , z) \nonumber \\
\label{eq:yy_hm3}
&& \times \left[
\int_{M_\mathrm{min}}^{M_\mathrm{max}} dM\, \frac{dn(M, z)}{dM}
b(M, z) y_\ell(M, z)
\right]^2 ,
\eeqa
where $k_\ell = \ell/\{ (1+z)d_A(z) \}$,
$z_\mathrm{dec}$ is the redshift of last scattering,
$d_A(z)$ is the angular diameter distance,
$d^2 V/dzd\Omega = (1+z)^2 d_A^2/H(z)$ is the comoving volume
per redshift and solid angle,
$y_\ell (M, z)$ is the Fourier transform of Compton-$y$ from
a single halo and
$P_\mathrm{m} (k, z)$ is the matter power spectrum.
The explicit formula of $y_\ell (M, z)$ is
\beq
\label{eq:y_ell}
y_\ell = \frac{4\pi R_s}{\ell_s^2} \frac{\sigma_\mathrm{T}}{m_\mathrm{e} c^2}
\int dx\, x^2 P_\mathrm{e}(x) \frac{\sin (\ell x/\ell_s)}{\ell x/\ell_s},
\eeq
where $x = r/R_s$, $\ell_s = d_A/R_s$, $R_s$ is the scale radius.
We define the halo radius $R_\Delta$ with the overdensity $\Delta$ as the radius
at which the mean density within $R_\Delta$ is equal to
$\Delta$ times to critical density $\rho_\mathrm{cr}$.
The enclosed mass $M_\Delta$ is defined as the mass within $R_\Delta$,
i.e.,
\beq
\label{eq:M_delta}
M_\Delta = \frac{4\pi}{3} \Delta \rho_\mathrm{cr}(z) R_\Delta^3 .
\eeq
For virial mass of halos $M_\vir$,
we use the expression from the top-hat collapse model
in \citet{Bryan1998},
\beq
\label{eq:Delta_vir}
\Delta_\vir = 18 \pi^2 +82(\Omega_\mathrm{m} (z)-1)
-39(\Omega_\mathrm{m}(z)-1)^2,
\eeq
where
\beq
\Omega_\mathrm{m} (z) = \Omega_\mathrm{m} (1+z)^3 E^{-2} (z) ,
\eeq
and
\beq
E(z) = H/H_0 =
[\Omega_\mathrm{m} (1+z)^3 + \Omega_\Lambda]^{1/2} .
\eeq
We adopt $M_\mathrm{200b}$, which is the enclosed mass within
the overdensity of $200$ times the mean background density,
as the halo mass $M$.
The corresponding overdensity is $\Delta = 200 \rho_\mathrm{m}(z) / \rho_\mathrm{cr} (z)
= 200 \Omega_\mathrm{m} (z)$.
The range of integration for halo mass is set as
$M_\mathrm{min} = 10^{12} \Msun/h$ and
$M_\mathrm{max} = 10^{16} \Msun/h$.
For halo mass function $dn(M, z)/dM$ and halo bias $b(M, z)$,
we adopt fitting formulae from \citet{Bocquet2016} and
\citet{Tinker2010}, respectively.
For convenience hereafter, we define $M_\mathrm{500c}$ as
the halo mass with the overdensity $\Delta = 500$.

\subsection{Cross-correlation of tSZ and WL}
\label{sec:cross}
Let us consider the cross-correlation of tSZ and WL.
The observable in WL observations which we focus on
is convergence field $\kappa (\bm{\theta})$.
The cross-power spectrum of Compton-$y$ and convergence
can also be computed based on the halo model prescription.
We can obtain the expression by replacing one of $y_\ell$
in Eqs.~\ref{eq:yy_hm2} and \ref{eq:yy_hm3} with $\kappa_\ell$,
which is the Fourier transform of the convergence signal from a single halo.
\beqa
\label{eq:yk_hm1}
C^{y\kappa} (\ell) &=& C^{y\kappa (\mathrm{1h})} (\ell) +
C^{y\kappa (\mathrm{2h})} (\ell), \\
C^{y\kappa (\mathrm{1h})} (\ell) &=& \int_0^{z_\mathrm{dec}} dz\,
\frac{d^2 V}{dzd\Omega} \nonumber \\
\label{eq:yk_hm2}
&& \times \int_{M_\mathrm{min}}^{M_\mathrm{max}} dM\, \frac{dn}{dM}
y_\ell (M, z) \kappa_\ell (M, z) , \\
C^{y\kappa (\mathrm{2h})} (\ell) &=& \int_0^{z_\mathrm{dec}} dz\,
\frac{d^2 V}{dzd\Omega} P_\mathrm{m} ( k_\ell, z) \nonumber \\
&& \times \int_{M_\mathrm{min}}^{M_\mathrm{max}} dM\, \frac{dn}{dM}
b(M, z) \kappa_\ell (M, z) \nonumber \\
\label{eq:yk_hm3}
&& \times \int_{M_\mathrm{min}}^{M_\mathrm{max}} dM\, \frac{dn}{dM}
b(M, z) y_\ell (M, z) .
\eeqa

Here, we briefly review how to compute the lensing signal from a single halo.
The density profile of dark halos is well described by
Navarro--Frenk--White (NFW) profile \citep{Navarro1996, Navarro1997},
\beq
\label{eq:rho}
\rho (r) = \frac{\rho_s}{(r/r_s) (1+r/r_s)^2} ,
\eeq
where $r_s$ is the scale radius and $\rho_s$ is the scale density.
The scale density $\rho_s$ is determined by the halo mass,
\beq
\label{eq:M_vir}
M_\vir = \int_0^{R_\vir} \rho (r) 4\pi r^2 \, dr =
4\pi \rho_s r_s^3 m_\mathrm{nfw} (c),
\eeq
where
\beq
m_\mathrm{nfw} (c) = \int_0^c \frac{x}{(1+x)^2}\, dx = \ln (1+c) - \frac{c}{1+c}.
\eeq
The parameter $c$ is the concentration parameter defined as $c = R_\vir/r_s$.
Throughout this paper, we adopt the following formula proposed by \citet{Duffy2008},
\beq
\label{eq:conc}
c(M_\vir, z) = 7.85 \left( \frac{M_\vir}{M_\mathrm{piv}} \right)^{-0.081} (1+z)^{-0.71} ,
\eeq
where $M_\mathrm{piv} = 2 \times 10^{12} \Msun/h$.
The halo model calculation needs the Fourier transform of the projected
density, i.e. convergence, denoted as $\kappa_\ell (M, z)$.
\beq
\label{eq:kappa_ell}
\kappa_\ell (M, z) =
\int 2\pi \theta \kappa (\theta) J_0 (\ell \theta) \, d\theta =
\frac{M\tilde{u}_M (k_\ell, z)}{d_A^2 \Sigma_\mathrm{crit} (z)} ,
\eeq
where $\kappa (\theta)$ is the convergence from a single halo,
$J_0 (x)$ is the zeroth-order Bessel function, $\tilde{u}_M (k)$
is the Fourier transform of $u_M (r)= \rho (r)/M$
and $\Sigma_\mathrm{crit} (z)$ is the critical surface mass density.
The analytical expressions of $\tilde{u}_M (k)$ and $\Sigma_\mathrm{crit} (z)$
are found in \citet{Oguri2011}.
For the calculation of $\Sigma_\mathrm{crit} (z)$ we need the redshift distribution
of source galaxies.
For RCSLenS, we adopt the following fitting function \citep{Harnois-Deraps2016},
\beqa
\label{eq:nz}
n_\mathrm{RCSLenS} (z) &=&
az \exp[-(z-b)^2/c^2] + dz \exp[-(z-e)^2/f^2] + \nonumber \\
& & gz \exp [-(z-h)^2/i^2],
\eeqa
where $(a, \allowbreak b, \allowbreak c, \allowbreak d, \allowbreak e,
\allowbreak f, \allowbreak g, \allowbreak h, \allowbreak i) =
(2.94,\allowbreak -0.44, \allowbreak 1.03, \allowbreak 1.58,
\allowbreak　0.40, \allowbreak 0.25, \allowbreak 0.38, \allowbreak　0.81,
\allowbreak 0.12)$.

In practice, two-point correlation function $\xi^{y \kappa} (\theta)$
is commonly used in observations.
We can transform the cross-power spectrum into the cross-correlation
via Hankel transformation,
\beq
\label{eq:xi_yk}
\xi^{y \kappa} (\theta) = \int \frac{\ell d \ell}{2 \pi} C^{y \kappa} (\ell) J_0 (\ell \theta) .
\eeq

\section{Methods}
\label{sec:methods}

\subsection{Semi-analytic model of the ICM}
\label{sec:semi}

In this section, we describe details of our model with $N$-body simulations.
Our model is semi-analytic in the sense that
the gas pressure profile for each halo is solved analytically
or adopted from the observed profile,
but the spatial distribution of halos are taken directly from $N$-body simulations.

First, we review the analytic gas profile briefly.
The model goes back to \citet{Ostriker2005}, and has been modified
in e.g., \citet{Shaw2010}, who introduced the concept of
radially dependent non-thermal pressure, and \citet{Flender2017},
who introduced a method for modeling cool cluster cores.

The main assumption in the model is that the gas re-arranges
inside the dark matter NFW profile into hydrostatic equilibrium
with a polytropic equation of state, which is described by the differential equation,
\beq
\label{eq:Euler_eq}
\frac{dP_{\mathrm{tot}}(r)}{dr} = -\rho_g(r) \frac{d\Phi(r)}{dr} ,
\eeq
where $P_{\mathrm{tot}}$ is the total (thermal + non-thermal) pressure,
$\rho_g$ is the gas density, and $\Phi$ is the dark matter NFW potential.
We can write the solution to this equation as,
\beqa
\label{eq:P_tot}
P_{\mathrm{tot}}(r) &=& P_0 \theta(r)^{n+1}, \\
\label{eq:rho_g}
\rho_{g}(r) &=& \rho_0 \theta(r)^{n} ,
\eeqa
where $\theta(r)$ is the polytropic variable,
\beq
\theta(r) = 1 + \frac{\Gamma-1}{\Gamma}\frac{\rho_0}{P_0}(\Phi_0 - \Phi(r)) ,
\eeq
and $\Phi_0$ is the central potential of the cluster.
Here, $\Gamma$ is the polytropic index, for which we adopt the value $1.2$,
in agreement with hydrodynamical simulations (e.g., \citealt{Nagai2007}).
In order to determine the shape of the NFW profile,
we measure the concentration parameter directly from the simulation
(for details, see Section~\ref{sec:simulation}).

Following \citet{Shaw2010}, we model the non-thermal pressure fraction as a power law,
\beq
\label{eq:nt_frac}
\frac{P_\mathrm{nt}}{P_\mathrm{tot}}(r) = \alpha(z)
\left( \frac{r}{R_{500}} \right)^{n_\mathrm{nt}} ,
\eeq
where $r$ is the distance from the center of halo, and the power law index
$n_\mathrm{nt}$ is a free parameter.
Since non-thermal pressure can not exceed total pressure,
at the outermost radius ($R_\mathrm{max}$), the inequality
$\alpha(z) \leq (R_\mathrm{max}/R_{500})^{-n_\mathrm{nt}}$
should be satisfied. Following \citet{Shaw2010}, we take the outermost radius
as $4 R_{500}$, and then it leads to $\alpha(z) \leq 4^{-n_\mathrm{nt}}$.
We parametrize the redshift dependent part as
\beq
\label{eq:alpha_z}
\alpha(z) = \alpha_0 \times \mathrm{min} [(1+z)^\beta, (f_\mathrm{max}-1)\tanh (\beta z) + 1],
\eeq
where $\alpha_0$ and $\beta$ are free parameters and
$f_\mathrm{max} = 4^{-n_\mathrm{nt}}/\alpha_0$.
Based on this functional form, at low redshift, the redshift dependence
is power law, but at high redshift, $f(z)$ asymptotes to the maximum value
$4^{-n_\mathrm{nt}}$. In our model, we fix $n_\mathrm{nt} = 0.8$ and $\beta = 0.5$
following \citet{Shaw2010} and constrain $\alpha_0$ with the power spectrum of tSZ
and the cross-correlation of tSZ and WL.
In addition, we keep $\alpha_0 < 4^{-n_\mathrm{nt}} = 0.33$ because
$\alpha_0$ greater than this value makes the pressure unphysical
($P_\mathrm{tot} < P_\mathrm{nt}$) at $r=R_{500}$.

We assume that a fraction of the gas mass has formed stars.
We model this fraction as a power-law,
\beq
\label{eq:shmr}
\frac{M_*}{M_\mathrm{500c}} = f_*
\left( \frac{M_\mathrm{500c}}{3 \times 10^{14} \Msun} \right)^{-S_*},
\eeq
where $M_*$ is the stellar mass,
$f_*$ is the stellar fraction at the pivot mass $3 \times 10^{14} \Msun$,
and $S_*$ is the mass-slope.

We further assume that some of the stars turn into supernovae and AGN,
which will induce feedback energy into the ICM given by $\epsilon_{\mathrm{f}} M_* c^2$,
with free parameter $\epsilon_{\mathrm{f}}$, which is typically small ($<10^{-5}$).
Another free parameter, $\epsilon_{\mathrm{DM}}$,
describes the amount of energy transfer from the dark matter
to the gas during major halo mergers via dynamical friction heating
(for a more detailed discussion, see \citet{Flender2017} and references therein).

In summary, six free parameters determine the ICM model,
[$\epsilon_\mathrm{DM}$, $\epsilon_\mathrm{f}$, $f_*$, $S_*$, $\alpha_0$, $\beta$].
In this analysis, we let the amount of non-thermal pressure, $\alpha_0$, vary,
and fix all other parameters to the best-fit values
from \citet{Flender2017},
$\epsilon_\mathrm{DM}=0.0$, $\epsilon_\mathrm{f}=3.97 \times 10^{-6}$,
$f_* = 0.026$, $S_* = 0.12$. We assume the fiducial value $\beta = 0.5$
adopted in \citet{Shaw2010}.

Alternatively, we also adopt the universal pressure profiles proposed
by \citet{Nagai2007} and calibrated using SZ observations
\citep{PlanckCollaboration2013},
\beqa
\label{eq:UPP}
\frac{P_\mathrm{e}(r)}{P_{500}} = p(x)
\left[ \frac{M_\mathrm{500c}}{3\times 10^{14} h_{70}^{-1} \Msun} \right]^{0.12}, \\
p(x) \equiv \frac{P_0}{(c_{500}x)^\gamma
[1+(c_\mathrm{500}x)^\alpha]^{(\beta-\gamma)/\alpha}} ,
\eeqa
where $(P_0, c_{500}, \gamma, \alpha, \beta) =
(6.41, 1.81, 0.31, 1.33, 4.13)$,
\beq
P_{500} = 1.65 \times 10^{-3} E(z)^{8/3}
\left[ \frac{M_\mathrm{500c}}{3\times 10^{14} h_{70}^{-1} \Msun} \right]^{2/3}
h_{70}^2 \,\mathrm{keV}\,\mathrm{cm}^{-3},
\eeq
$x = r/R_{500}$ and $h_{70} = h/0.7$.
Note that the sample used in calibration consists of clusters
of which the mass range is from $0.9$ to $15 \times 10^{14} \Msun$ and
the redshift is less than $0.5$.
While Eq.~\ref{eq:UPP} can reproduce the pressure profile for halos at this range,
the pressure profile of group size halos and high redshift halos still remain uncertain.
The above fitting formula assumes hydrostatic equilibrium,
which leads to the bias of the mass estimate.
Following \citet{Dolag2016}, we rescale $M_\mathrm{500c} \to M_\mathrm{500c}/(1+b_\mathrm{HSE})$
and $R_{500} \to R_{500}/(1+b_\mathrm{HSE})^{1/3}$,
where $b_\mathrm{HSE}$ is the hydrostatic bias and we adopt $b_\mathrm{HSE} = 0.2$
as the fiducial value.
We use the analytic profile and the universal pressure profile to
constrain $\sigma_8$ and non-thermal pressure amplitude
in Section~\ref{sec:constraint}.

\subsection{Numerical simulations and map making procedure}
\label{sec:simulation}
\begin{figure}
\includegraphics[width=8cm]{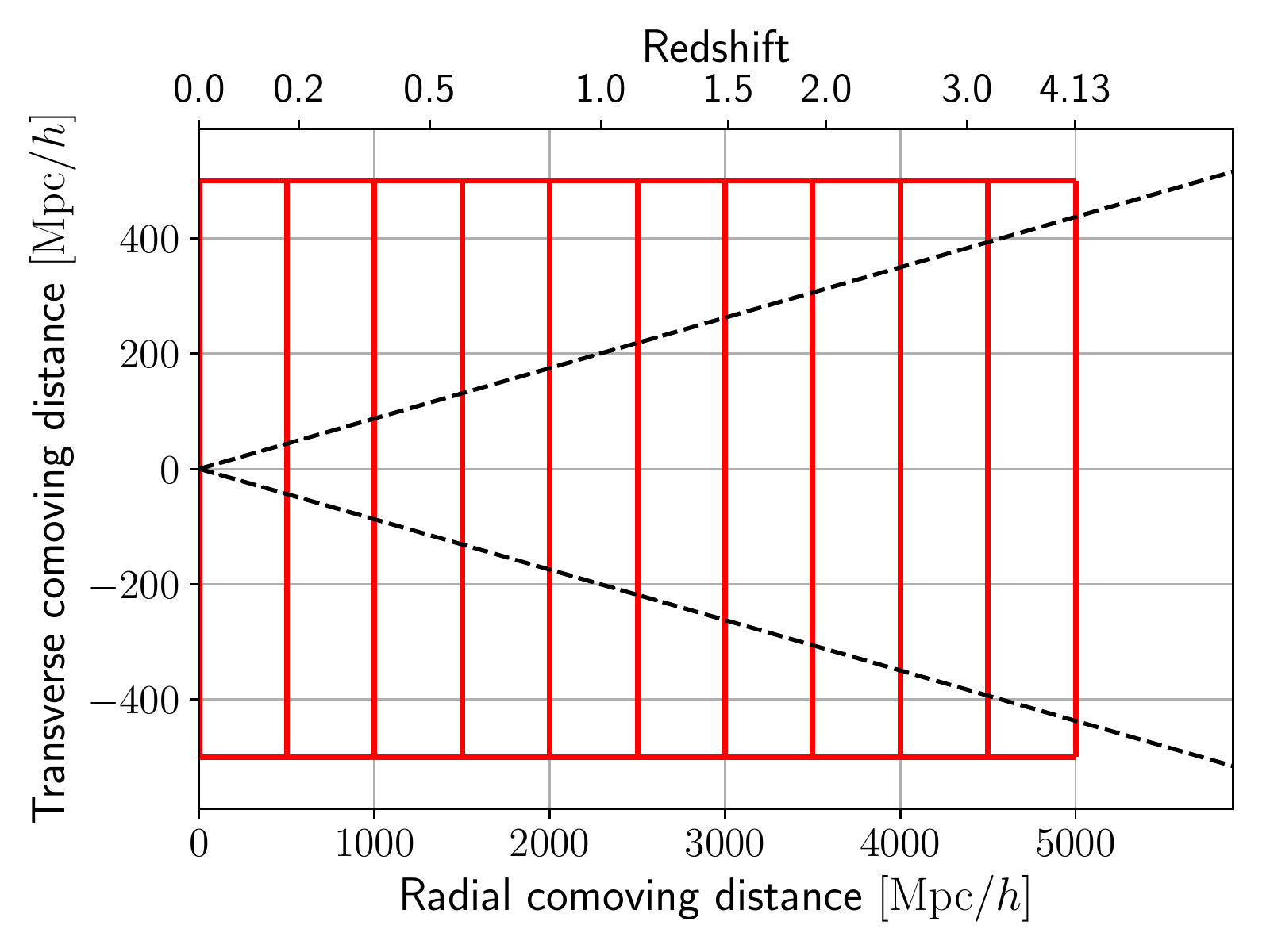}
\caption{Configuration of snapshots. Each red box corresponds to
a single snapshot and the dashed lines show the extent of the light-cone.
The length in the line-of-sight direction is $500\,\mathrm{Mpc}/h$
and we randomly extract the corresponding region from the original snapshot,
which has $1000\,\mathrm{Mpc}/h$ on a side.
The dashed lines show the opening angle with $10\,\mathrm{deg}$.}
\label{fig:design}
\end{figure}

First, we run an $N$-body simulation to obtain the spatial distribution of matter
in the Universe at different redshifts.
We use Tree-PM code {\tt Gadget-2} \citep{Springel2005}.
The number of particles is $2048^3$,
the volume of the simulation box is $(1\,\mathrm{Gpc}/h)^3$, and
the corresponding particle mass is $m_\mathrm{p} = 1.02 \times 10^{10} \Msun /h$.
We generate the initial condition at the redshift $z_\mathrm{ini} = 59$
with a parallel code developed in
\citet{Nishimichi2009, Nishimichi2010, Valageas2011},
which employs second order Lagrangian perturbation theory.
We store 10 snapshots to construct a light-cone output from $z=4.13$ to $z=0.0$.
The redshifts at which snapshots are stored are determined to satisfy
$\chi(z_{i+1}) - \chi(z_i) = 500\,\mathrm{Mpc}/h\ (i=1,\ldots,9)$ and
$\chi(z_1) = 250\,\mathrm{Mpc}/h$
(see Figure~\ref{fig:design}).
For halo finding, we employ the {\tt Rockstar} halo finder \citep{Behroozi2013}.
We assign gas pressure to each particle which belongs to any halo
according to the radius from the center
based on the analytic model presented in Section \ref{sec:semi}.
If a particle does not belong to any halos, it does not contribute to tSZ signal.
Since the code automatically provides the concentration parameter
by fitting the density profile with NFW profile,
we use this concentration parameter instead of the fitting formula
of concentration parameters.

In order to carry out mock observations for WL,
we employ the multiple-plane ray-tracing method
\citep{White2000, Hamana2001, Sato2009, Hilbert2009}.
First, we place snapshots to create the light-cone
which fills the volume from $z=0$ to $z=4.13$.
For each snapshot, we pick $500\,\mathrm{Mpc}/h$ slice,
half of the simulation box, in the line-of-sight direction
and then randomly rotate and translate particles keeping periodic boundary condition
so that the same structure does not appear multiple times.
The angular extent of each map is $10\degr \times 10\degr$
and the number of grids on a side is $8192$, which corresponds to the pixel
size of $10\degr / 8192 \simeq 0.073\, \mathrm{arcmin}$.
Finally, by repeating the random rotation and translation 100 times,
we generate 100 mock $10\degr \times 10\degr$ convergence maps,
applying weights derived from the source redshift distribution
(Eq.~\ref{eq:nz}).

\begin{figure*}
\includegraphics[width=14cm]{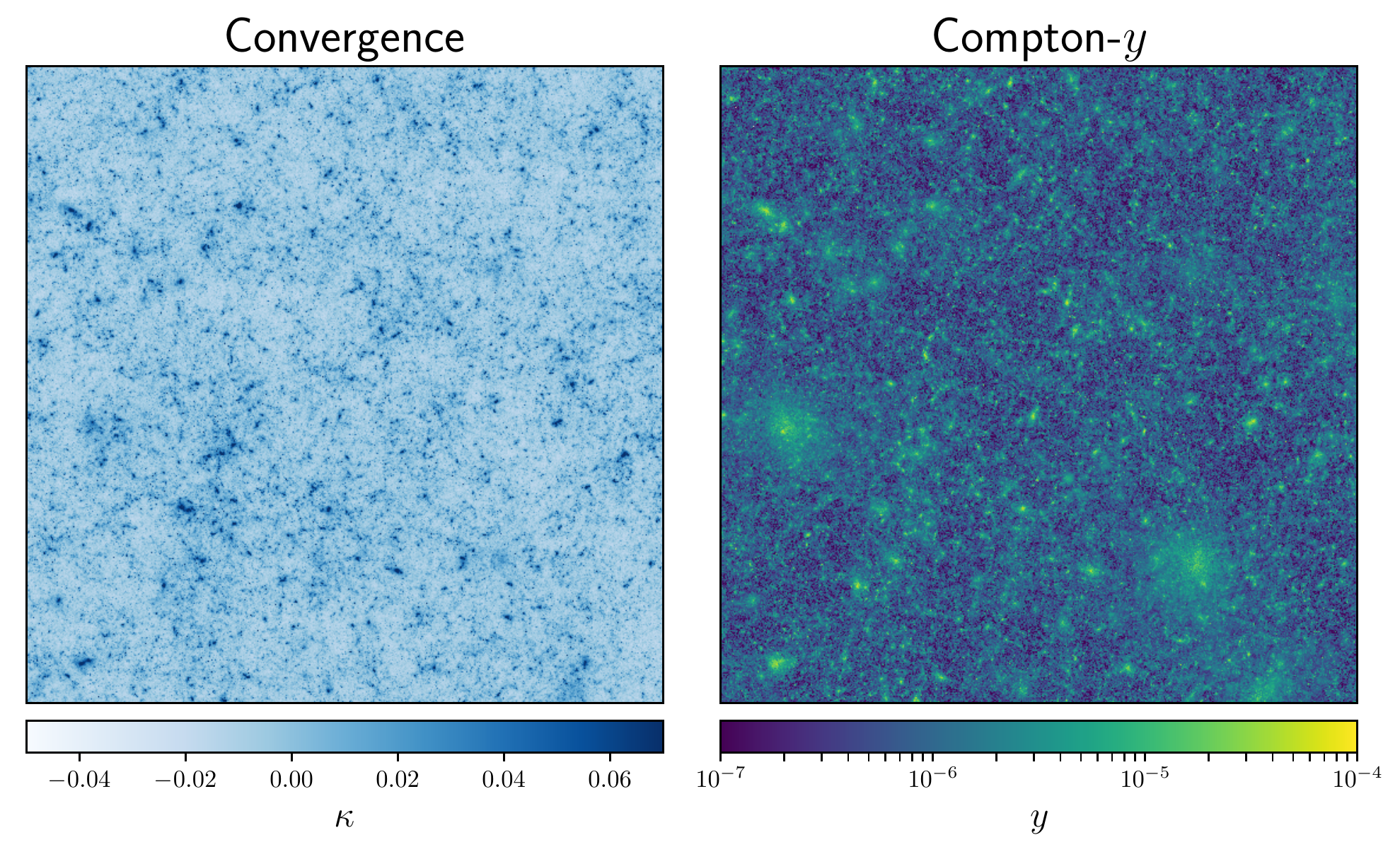}
\caption{Convergence and Compton-$y$ maps obtained from simulations.
The length on a side corresponds to $10\degr$.}
\label{fig:map_images}
\end{figure*}

Similarly, we create mock Compton-$y$ maps based on the method presented in
\citet{Roncarelli2007, Ursino2010}.
For the Compton-$y$ map, we do not include ray deflection effect because
the effect is negligible at the scales where measurements are available
\citep{Troster2014}.
For sanity check, we measure
the average Compton-$y$ parameter $\langle y \rangle$.
For the semi-analytic pressure profile,
$\langle y \rangle = (1.47 \pm 0.10) \times 10^{-6}$
and for the universal pressure profile,
$\langle y \rangle = (1.07 \pm 0.10) \times 10^{-6}$.
The error corresponds to the standard deviation over 100 mock maps.
The results are close to that of the previous study \citep{Dolag2016}
based on hydrodynamics simulations,
$\langle y \rangle = 1.18 \times 10^{-6}$.
Note that they adopted the different hydrodynamics model
and cosmological parameters.
For reference, \citet{Khatri2015} presented bounds of the average
Compton-$y$ from which they subtracted the contribution from galaxy clusters
as $5.4 \times 10^{-8} < \langle y \rangle < 2.2 \times 10^{-6}$.
Figure~\ref{fig:map_images} shows ones of convergence and Compton-$y$ maps
as an example.

In order to make our simulated maps more realistic,
we add noise to convergence and Compton-$y$ maps,
and then smooth them with the Gaussian filter.
For tSZ maps, following \citet{Dolag2016},
we add the Gaussian noise so that the standard deviation of the noise map
is $\sigma = 1.5 \times 10^{-6}$ with the FWHM window scale
$\theta_\mathrm{FWHM} = 10\,\mathrm{arcmin}$.
For weak lensing, the dominant source of the noise is the shape noise
and the noise can be modeled as Gaussian \citep{VanWaerbeke2000}.
The variance of the noise is given as,
\beq
\sigma^2 = \frac{\sigma^2_\epsilon}{\theta^2_\mathrm{pix} n_\mathrm{gal}},
\eeq
where $\sigma_\epsilon$ is the standard deviation of the intrinsic
ellipticity, $\theta_\mathrm{pix}$ is the pixel size of the map
and the $n_\mathrm{gal}$ is the mean number density of the source galaxy.
For RCSLenS, we adopt $\sigma_\epsilon = 0.277$ and
$n_\mathrm{gal} = 5.8\,\mathrm{arcmin}^{-2}$ \citep{Hojjati2017}.
After adding the noise, both of the maps are smoothed with the Gaussian
filter with $\theta_\mathrm{FWHM} = 10\,\mathrm{arcmin}$
which is the same smoothing scale in creating Compton-$y$ map
\citep{Planck2015tSZ}.

\subsection{Estimation of the covariance matrix}
\begin{figure}
\includegraphics[width=8cm]{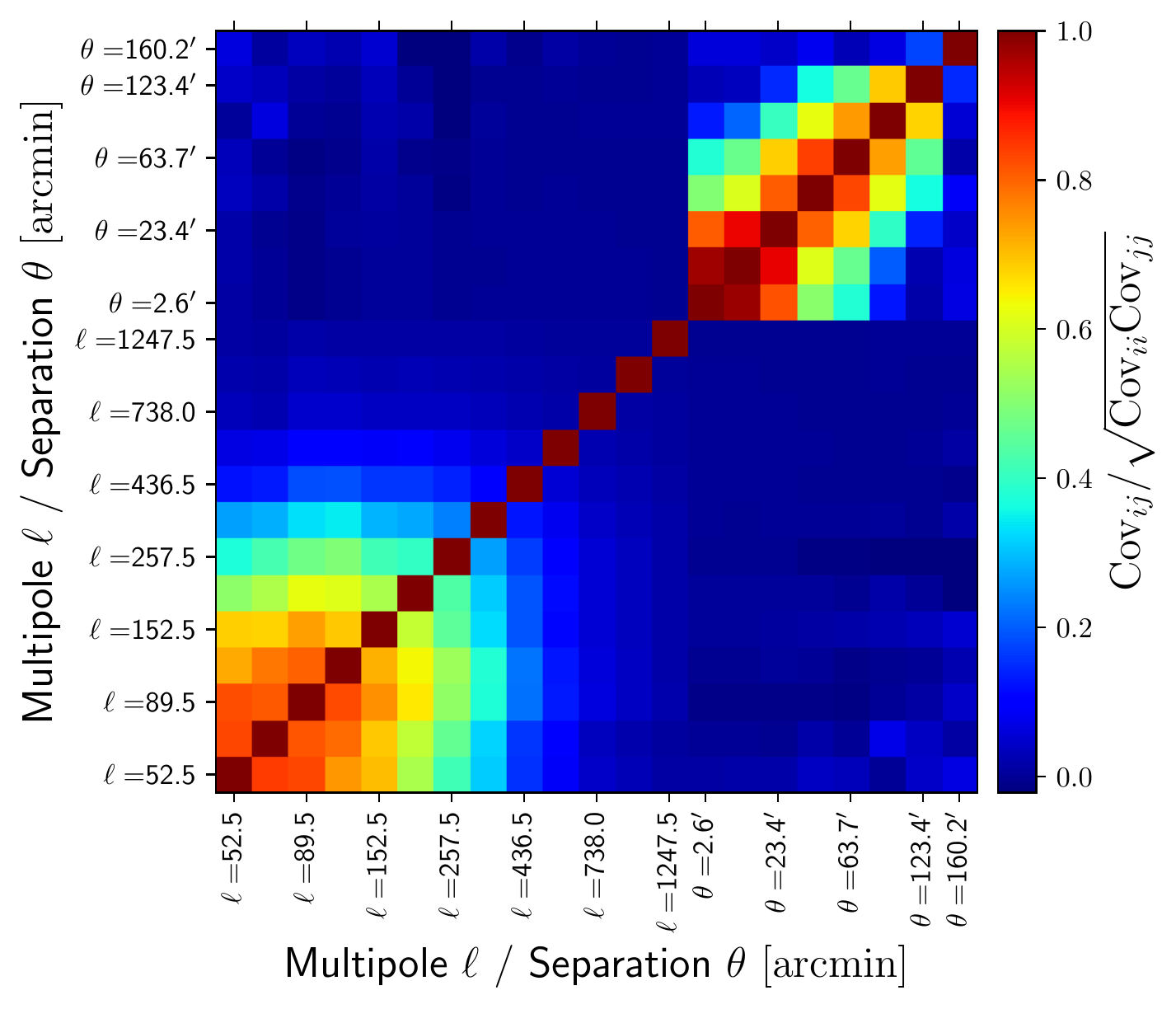}
\caption{Covariance matrices measured from simulations and Gaussian maps.
From 1st to 13th rows and columns correspond to the power spectra, and
from 14th to 21st rows and columns do to cross-correlations.
The upper left (lower left) part corresponds to the covariance
with the analytic (observed) pressure profile.
The scaling factor due to the size of areas has already been applied.}
\label{fig:cov}
\end{figure}

We present how we measure the covariance matrix from the mock maps
generated from simulations.
In our analysis, the data vector $\bm{N}$ is defined as,
\beq
\bm{N} = (C^{yy} (\ell_1), \ldots, C^{yy} (\ell_{n_C}),
\xi^{y \kappa} (\theta_1), \ldots, \xi^{y \kappa} (\theta_{n_\xi}))^{\mathrm{T}} ,
\eeq
where the dimensions of the data vectors are
$n_C = 13\,(52.5 \leq \ell \leq 1247.5)$ and
$n_\xi = 8\,(2.55 \leq \theta/\mathrm{arcmin} \leq 160)$.
Though in {\it Planck} data, there are more available data points for lower multipoles,
we do not use these data points due to the size of mock maps.
We have 100 mock maps and as a result $R=100$ measurement of the data vector
$\bm{N}^{r}\,(r=1,\ldots,R)$.
The area of mock maps is $100\,\mathrm{deg}^2$, but we will apply this covariance
matrix to the measurements by {\it Planck} and RCSLenS,
both of which have larger survey areas.
We need to scale the covariance matrix according to the survey area.
The estimated covariance matrix is expressed as,
\beq
\mathrm{Cov}_{ij} =
f^s_{ij} \frac{1}{R-1} \sum_{r=1}^R (N_i^r-\bar{N}_i) (N_j^r-\bar{N}_j),
\eeq
where $\bar{N}$ is the mean over the $R$ realizations,
\beq
\bar{\bm{N}} = \frac{1}{R} \sum_{r=1}^R \bm{N}^r ,
\eeq
and $f^s_{ij}$ is the scaling factor of the survey area,
\beq
f^s_{ij} =
\begin{cases}
  A_\mathrm{sim}/A_\mathrm{Planck} & (\text{between the power spectrum}) \\
  A_\mathrm{sim}/A_\mathrm{RCSLenS} & (\text{between the cross-correlation}) .
\end{cases}
\eeq
The survey areas are $A_\mathrm{sim} = 100\,\mathrm{deg}^2$,
$A_\mathrm{Planck} = 20626\,\mathrm{deg}^2$ and
$A_\mathrm{RCSLenS} = 560\,\mathrm{deg}^2$.
For covariance between the power spectrum and the cross-correlation,
there is no appropriate scaling factor because the sizes of the survey areas
of {\it Planck} and RCSLenS are different.
In order to estimate the cross-covariance, we generate 100 Gaussian maps
of Compton-$y$ and convergence which reproduce
the power spectrum $C^{yy}(\ell)$ and
cross-spectra $C^{y\kappa} (\ell)$ computed from the halo model
with the fiducial parameters.
The size of Gaussian maps is matched with the survey area
of {\it Planck} (RCSLenS) for Compton-$y$ (convergence) maps.
Then, we compute the power spectra and the cross-correlations
based on these maps, and estimate the cross-covariance
as the variance over 100 Gaussian maps.
For the power spectrum of tSZ, we take into account
the variance due to incomplete separation
between the signal of tSZ and contaminants, e.g., cosmic infrared background.
In order to estimate the variance, we use the values reported
by \citet{Planck2015tSZ}.
In Figure~\ref{fig:cov}, the covariance matrices measured
from our simulations and Gaussian maps are shown.
For the power spectrum part, though mainly the diagonal components are dominated,
there are substantial off-diagonal correlations caused by
the connected trispectrum term \citep{Horowitz2017}.

\section{Results}
\label{sec:results}

\subsection{Power spectrum and cross-correlation}
\label{sec:power_cross}
\begin{figure}
\includegraphics[width=8cm]{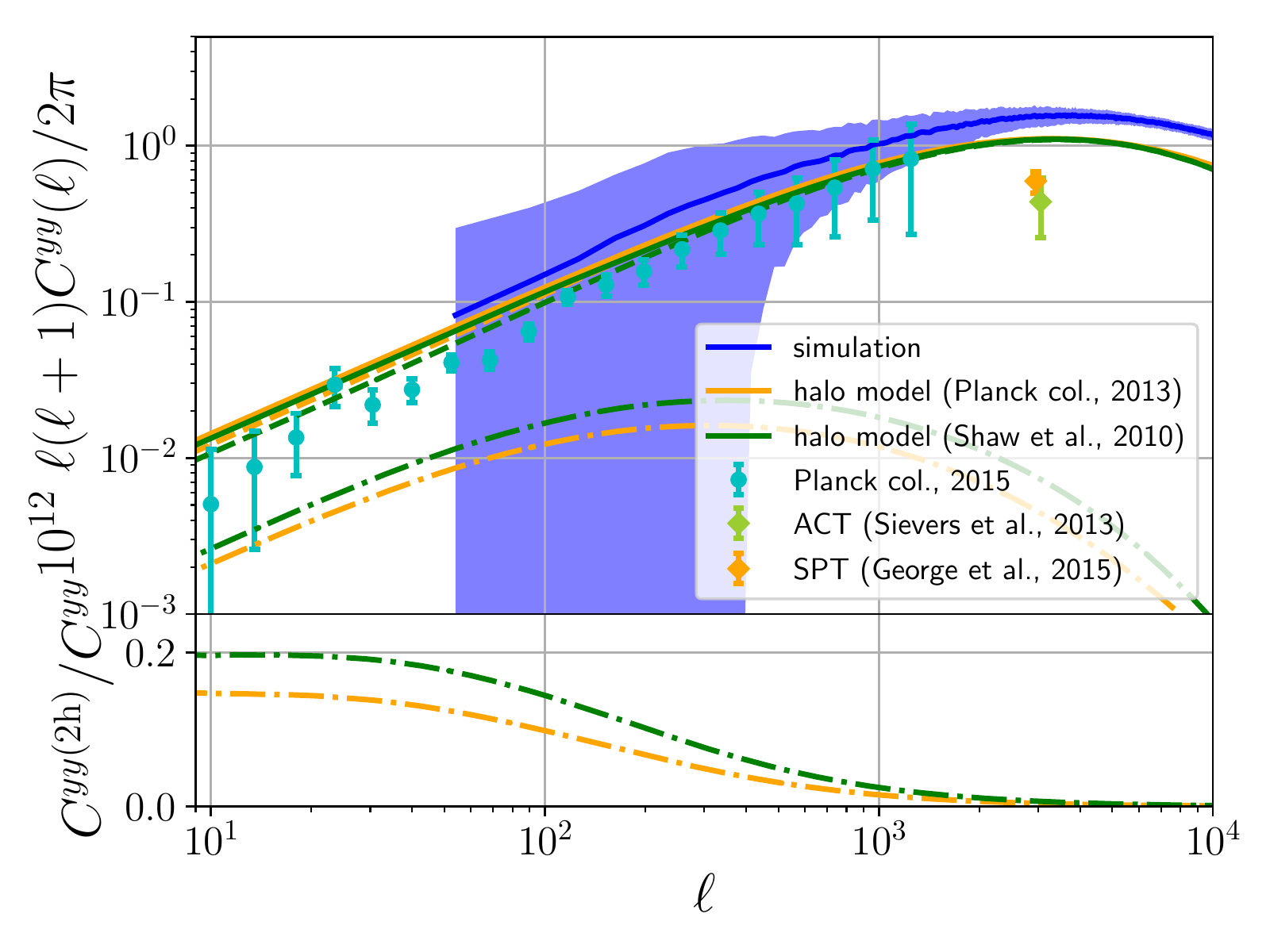}
\caption{Power spectra of Compton-$y$ from simulations, the halo model prediction
of two different pressure profile.
The dashed (dot-dashed) line shows 1-halo (2-halo) contribution.
The lower panel shows the ratio of 2-halo term to the total spectrum.
For comparison, the observational estimates from {\it Planck} \citep{Planck2015tSZ},
ACT \citep{Sievers2013}, and SPT \citep{George2015}.
are also shown.
The shaded region corresponds to the standard deviation over 100 mock maps which
cover $100$ square degrees.}
\label{fig:yy}
\end{figure}
We show power spectra of Compton-$y$ for different models
in Figure~\ref{fig:yy}.
The results of the analytic model \citep{Shaw2010} and the simulation based
semi-analytic model are not consistent at smaller scales ($\ell \simgt 2000$)
possibly due to the lack of resolution in $N$-body simulations.
In addition, the effects of the asphericity and substructures
can explain part of the differences \citep{Battaglia2012}.
However, at scales which can be accessible by {\it Planck} data
($100 \simlt \ell \simlt 1000$), both models give consistent results.
For even larger scales ($\ell \simlt 100$), the power spectra of the semi-analytic model
is suppressed and the variance is quite large affected by the size of mock maps.
Overall, all of the results overestimate the power spectrum compared with
the measurement of {\it Planck}. One of the possible reasons is that
our input parameter $\sigma_8 = 0.831$ is high. We will address this point
in the following Section.

\begin{figure}
\includegraphics[width=8cm]{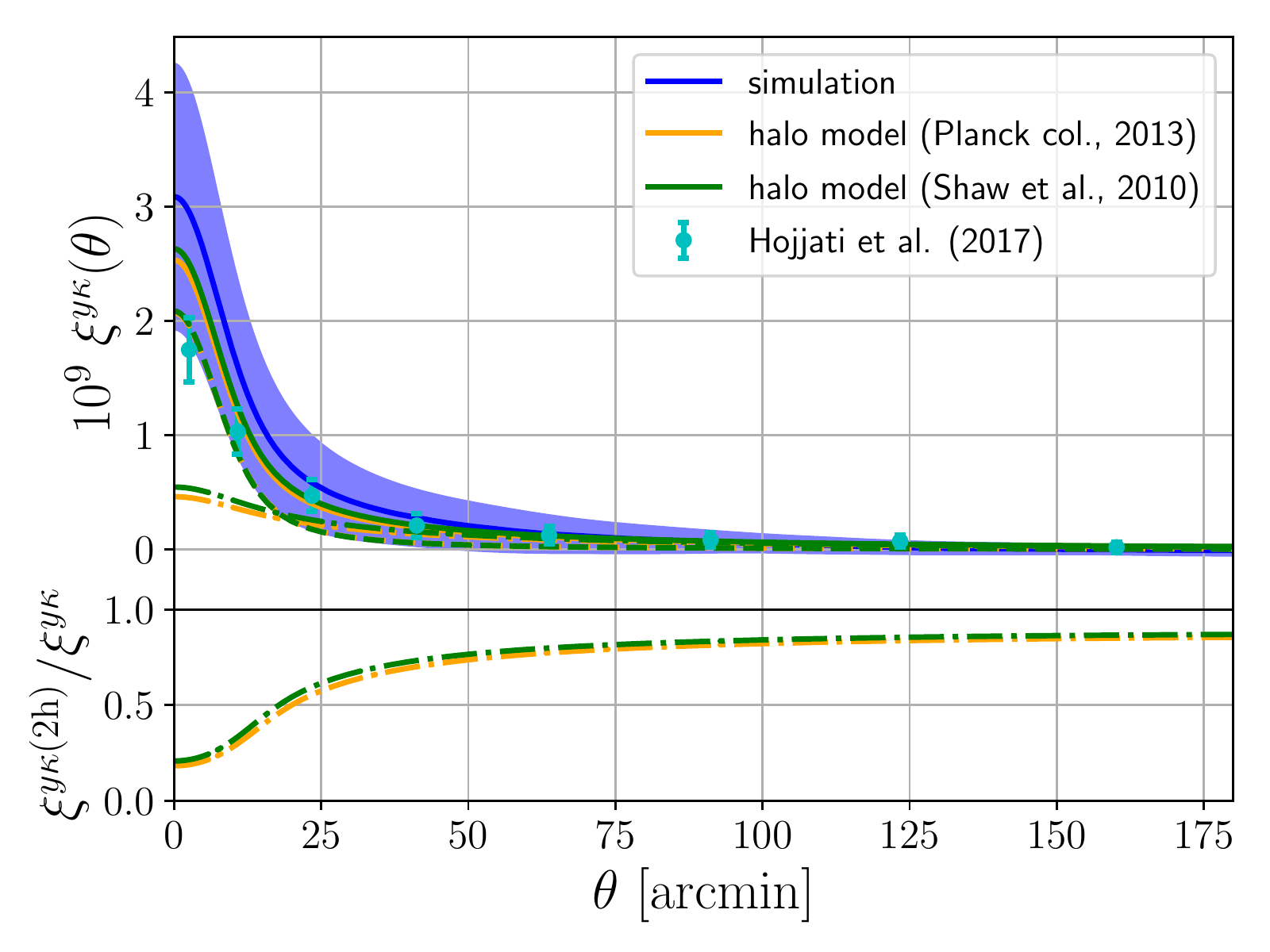}
\caption{Cross-correlation function of tSZ and WL from our simulation and
halo model calculations.
The dashed (dot-dashed) line shows 1-halo (2-halo) contribution.
The lower panel shows the ratio of 2-halo term to the total cross-correlation.
The shaded region corresponds to the standard deviation
over 100 mock maps which cover $100$ square degrees.}
\label{fig:yk}
\end{figure}
Figure~\ref{fig:yk} shows the cross-correlation of tSZ and WL
from our simulation based semi-analytic model and halo model calculations.
Although the excess of the cross-correlations at small scales
($\theta < 10\,\mathrm{arcmin}$) can be seen as a possible tension,
the results are consistent with each other on larger scales. This difference
also can be induced by the high value of $\sigma_8$.

\subsection{Constraints on non-thermal pressure and $\sigma_8$}
\label{sec:constraint}
With the power spectrum and the cross-correlation
measured by {\it Planck} and RCSLenS, we can constrain
the amplitude $\alpha_0$ of the non-thermal pressure and the amplitude
of the matter power spectrum, i.e., $\sigma_8$.
Other model parameters and cosmological parameters are fixed
at the fiducial values.
The posterior distribution when both of the power spectrum and
the cross-correlation are used is given as,
\beqa
\log L(\alpha_0, \sigma_8 | \bm{N}_\mathrm{data}) =
-\frac{1}{2} \log [ (2\pi)^{n_C+n_\xi}|\det \mathrm{Cov}| ] \nonumber \\
-\frac{1}{2}(\bm{N}_\mathrm{data}-\bm{N}_\mathrm{model})^{\mathrm{T}}
(\mathrm{Cov})^{-1} (\bm{N}_\mathrm{data}-\bm{N}_\mathrm{model}) ,
\eeqa
where $\bm{N}_\mathrm{data} = (\bm{C}^{yy}_\mathrm{Planck}, \bm{\xi}^{y\kappa}_\mathrm{RCSLenS})$
and $\bm{N}_\mathrm{model}$ is the halo model prediction given $\alpha_0$ and $\sigma_8$.
When we use either the power spectrum or the cross-correlation,
we simply use a submatrix of the covariance and a subvector of the model vector.

\begin{figure}
\includegraphics[width=8cm]{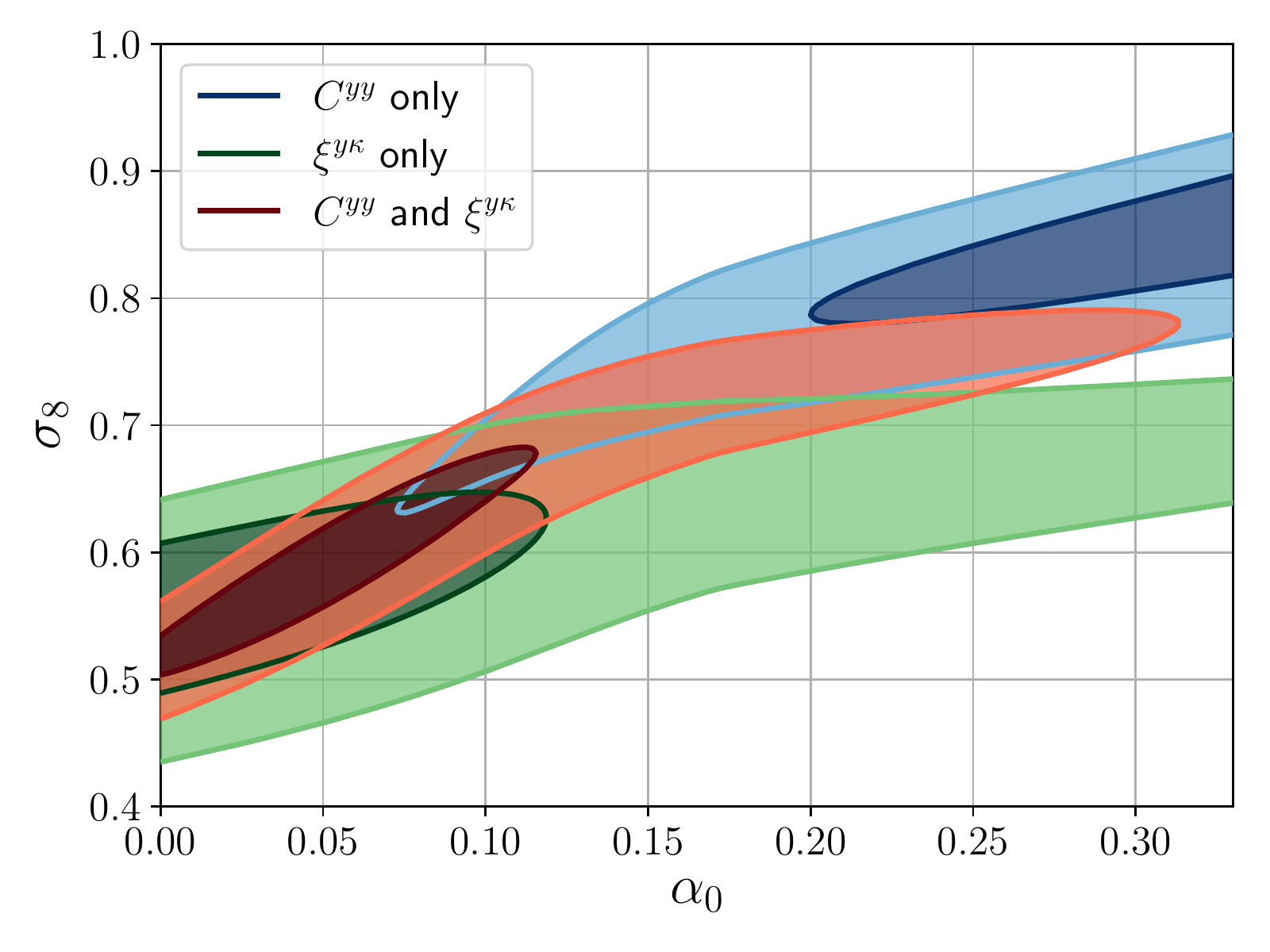}
\caption{Posterior distributions of non-thermal pressure parameters $\alpha_0$
and $\sigma_8$.
The inner (outer) colored region correponds to $1\sigma$ ($2\sigma$)
confidence level.
The results with data sets of power spectra and cross-correlations,
power spectra only, and cross-correlations are shown in solid red, blue, and green
regions, respectively.}
\label{fig:cr}
\end{figure}

We estimate the probability contours by computing the posterior probability
at regular grids.
The posterior distribution is shown in Figure~\ref{fig:cr} with different data sets,
power spectrum only, cross-correlation only, both of them.
The red, blue, and green solid lines correspond to the confidence regions
with the data sets of both of power spectra and cross-correlations, power spectra only,
and cross-correlations only, respectively.
With all data sets, the clear degeneracy between $\sigma_8$ and $\alpha_0$ can be seen.
If only the power spectra are employed, moderate $\sigma_8 \sim 0.85$
are preferred but the estimated $\alpha_0$ is clearly larger than
the fiducial value $0.18$.
On the other hand, the results with cross-correlations (red and green lines) prefer
low $\sigma_8 \sim 0.6$ and low $\alpha_0 \sim 0.05$.
The low non-thermal pressure amplitude $\alpha_0$
is strongly inconsistent with the predictions based on hydrodynamical
cosmological simulations \citep{Shaw2010, Nelson2014}.
The estimated value of $\sigma_8$ is quite smaller than the result
from CMB measurements of {\it Planck}, $\sigma_8 = 0.831 \pm 0.013$
\citep[TT,TE,EE+lowP,][]{Planck2015parameters}.
However, recent analysis of KiDS weak lensing survey \citep{Kohlinger2017}
reports $\sigma_8 \sqrt{\Omega_\mathrm{m}/0.3} = 0.651 \pm 0.058$, i.e.
$\sigma_8 = 0.635 \pm 0.057$ for $\Omega_\mathrm{m} = 0.3156$, which is consistent
with our result within $1\sigma$ level.

\begin{figure}
\includegraphics[width=8cm]{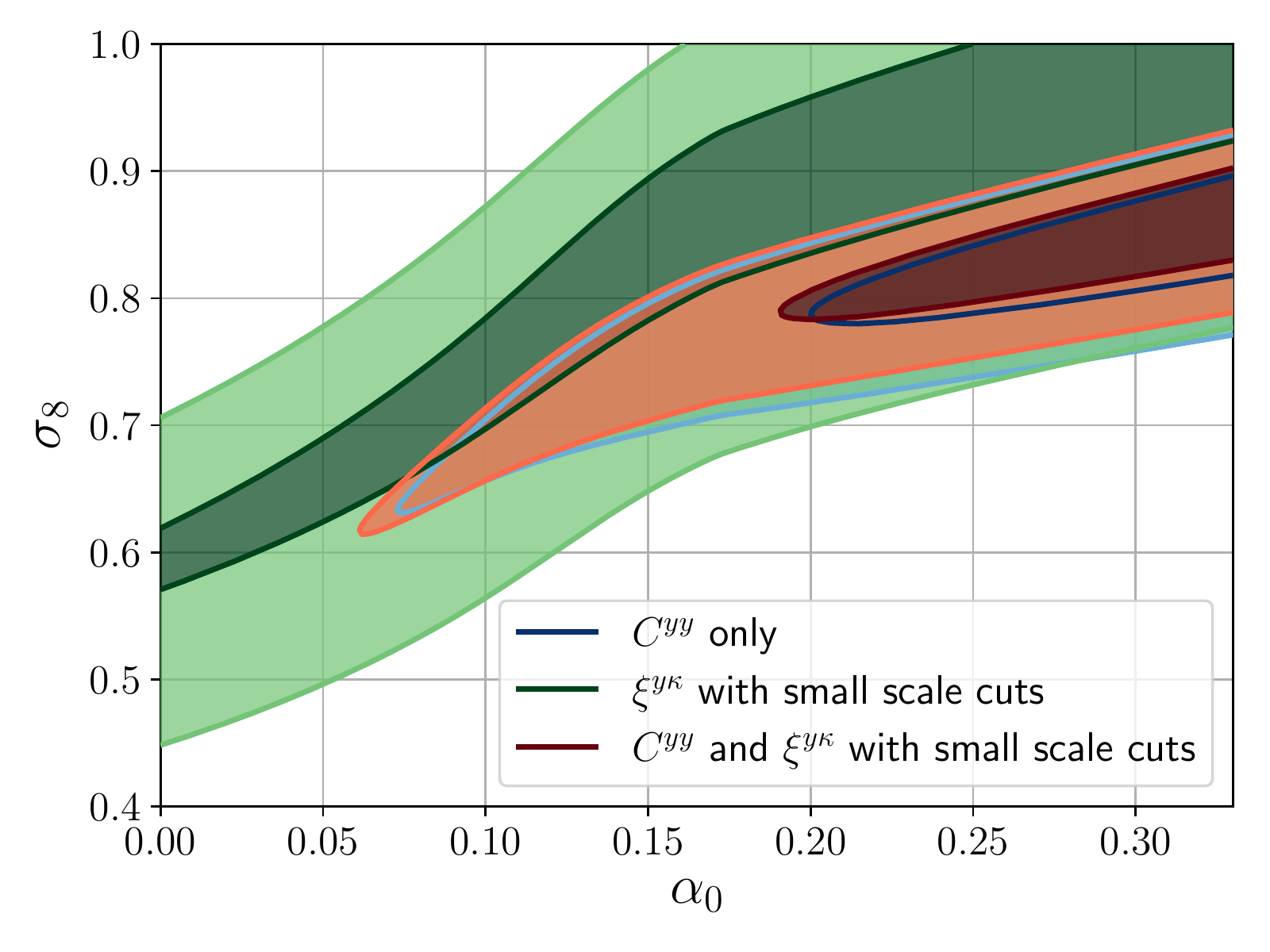}
\caption{Posterior distributions of non-thermal pressure parameters $\alpha_0$
and $\sigma_8$ with small scale ($< 10\, \mathrm{arcmin}$) cuts in cross-correlations.
The inner (outer) colored region correponds to $1\sigma$ ($2\sigma$)
confidence level.
The results with data sets of power spectra and cross-correlations,
power spectra only, and cross-correlations only are shown in solid red, blue, and green
regions, respectively. Note that the blue regions are identical in Figure~\ref{fig:cr}
because this data set does not include cross-correlations.}
\label{fig:cr_cut}
\end{figure}

In addition, we investigate the effect of the small scale (less than $10\,\mathrm{arcmin}$,
which is the smoothing scale)
cross-correlations. Figure~\ref{fig:cr_cut} shows the confidence regions with
small scale cross-correlations excluded.
In these cases, all of results become consistent with each other.
This result indicates that the tension originates from the small scales.

\begin{figure}
\includegraphics[width=8cm]{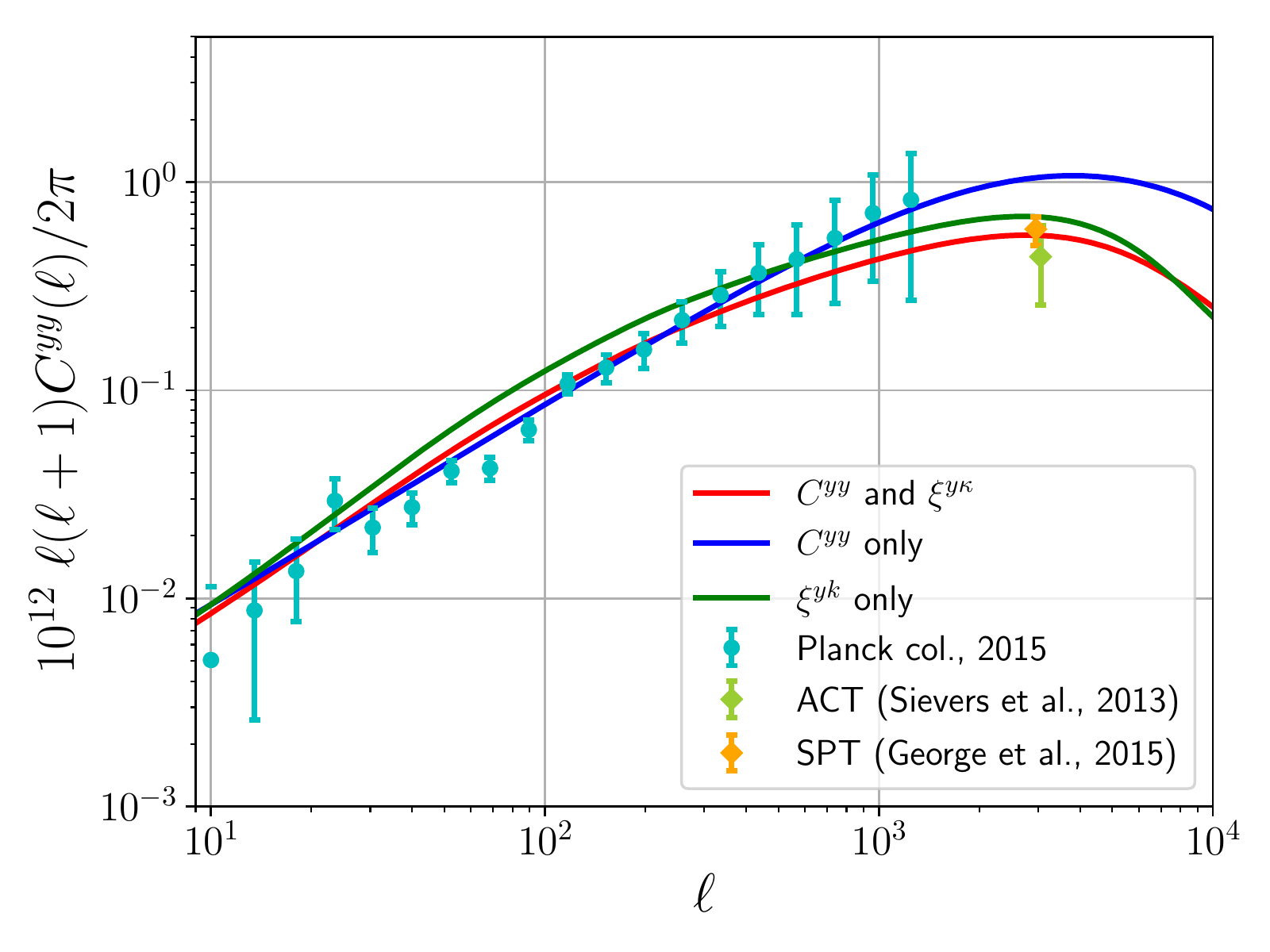}
\includegraphics[width=8cm]{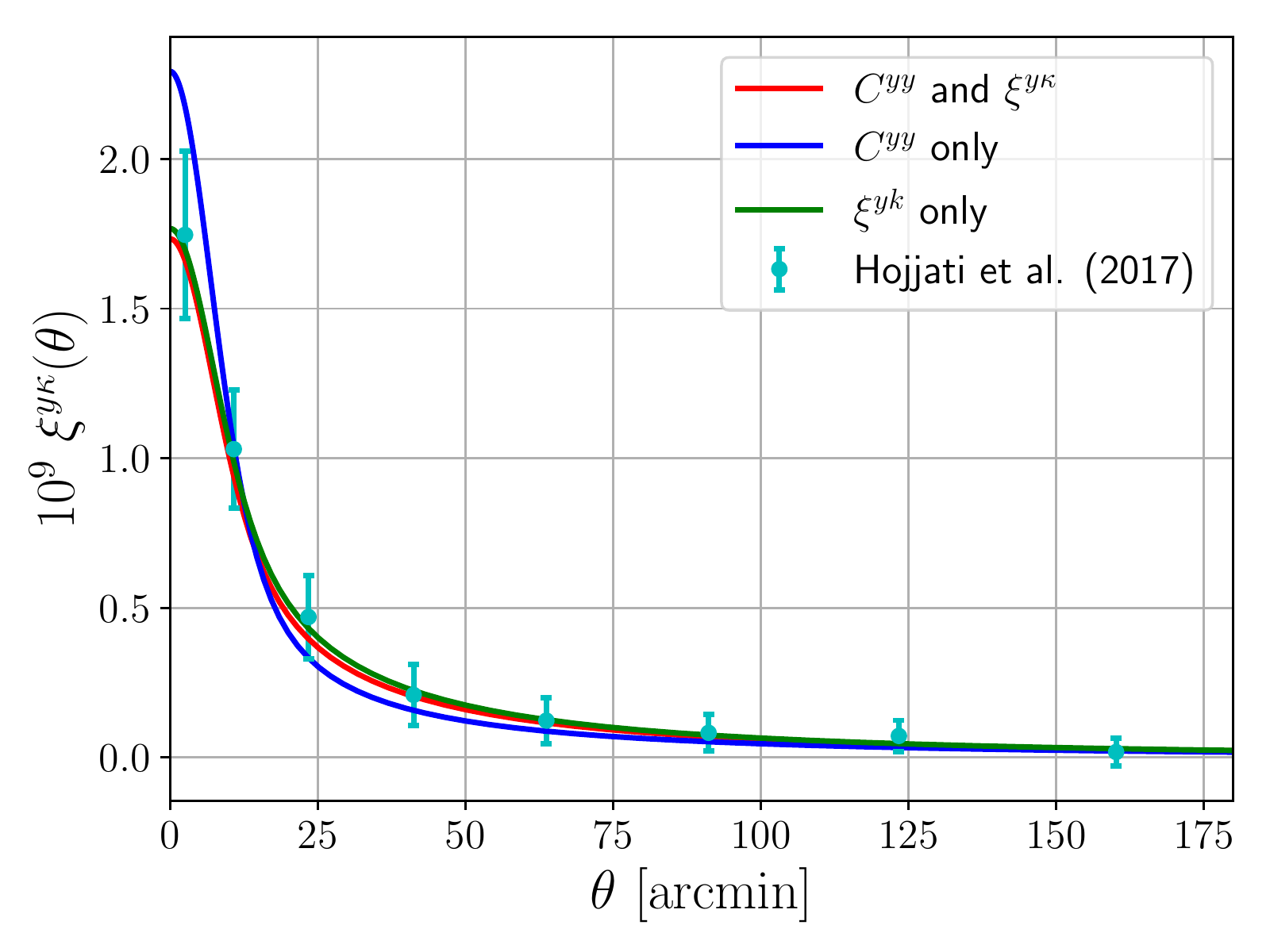}
\caption{Power spectra and cross-correlations with best-fit parameters.
The results using data sets of power spectra only, cross-correlations only, and both
are shown as blue, green, and red lines, respectively.}
\label{fig:bf}
\end{figure}

In Figure~\ref{fig:bf} we show the tSZ power spectrum and tSZ-WL cross-correlation,
together with the best-fit model parameters estimated with
the data sets of power spectrum only, cross-correlation only, and both.
Remarkably, when we include cross-correlations, the best-fit power spectrum
can reproduce ACT and SPT data points, though these data points are not used
in the analysis.

\begin{figure}
\includegraphics[width=8cm]{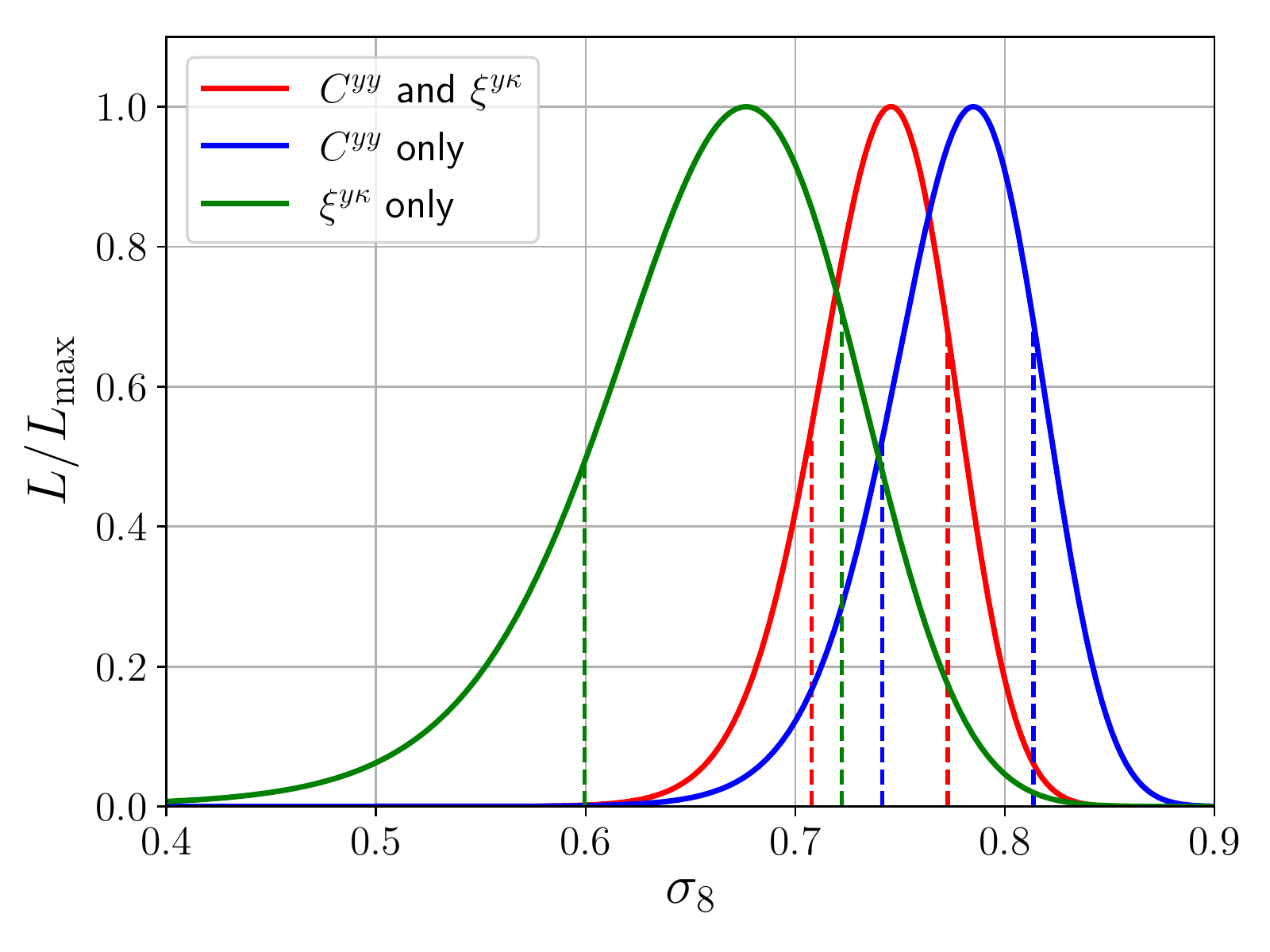}
\caption{Posterior distributions of $\sigma_8$ with the power spectrum
and the cross-correlation using the observationally calibrated universal pressure profile.
The dashed lines show the $16\%$ and $84\%$ percentile.}
\label{fig:sigma8}
\end{figure}

We next derive constraints on $\sigma_8$ using
the universal pressure profile with parameters
calibrated against {\it Planck} data (Eq.~\ref{eq:UPP}).
Note that we apply the pressure profile to less massive and/or
high redshift halos, which are not calibrated in this pressure profile.
Using only the tSZ power spectrum data,
we find $\sigma_8 = 0.785^{+0.029}_{-0.043}$,
consistent with \citet{Planck2015tSZ},
who find $\sigma_8 (\Omega_\mathrm{m}/0.28)^{3/8} = 0.80^{+0.01}_{-0.03}$, i.e.,
$\sigma_8 = 0.76^{+0.01}_{-0.03}$ for $\Omega_\mathrm{m} = 0.3156$,
from a similar analysis.
The tSZ-WL cross-correlation on the other hand
prefers a lower value of $\sigma_8 = 0.677^{+0.046}_{-0.077}$.
Combining the two data sets, we find $\sigma_8 = 0.746^{+0.026}_{-0.038}$.
The posterior distributions derived from tSZ and the tSZ-WL cross-correlation
show a clear tension (see Fig.~\ref{fig:sigma8}).

\begin{table}
\caption{Summary of constraints on $\sigma_8$.
The best-fit value and $16\%$ and $84\%$ percentile values are shown.}
\label{tab:constraints}
\begin{tabular}{lc}
\hline
Data sets & Constraints of $\sigma_8$\\
\hline
$C^{yy}$ and $\xi^{y\kappa}$ & $0.746^{+0.026}_{-0.038}$\\
$C^{yy}$ only & $0.785^{+0.029}_{-0.043}$\\
$\xi^{y\kappa}$ only & $0.677^{+0.046}_{-0.077}$\\
\hline
\end{tabular}
\end{table}

\subsection{Mitigating the tension between the data sets}
As seen above, the constraints on $\sigma_8$ and $\alpha_0$
from the tSZ power spectrum and the tSZ-WL cross-correlation are inconsistent.
The tension seems to originate from the small scales,
as we have seen in Figure~\ref{fig:cr_cut}.
Here, we investigate if modifications to the gas model can help mitigate the tension.
The analytic pressure profile is calibrated against X-ray observations of
massive clusters over a wide range of redshift, and
low redshift galaxy groups \citep{Flender2017}.
Therefore, the gas profile of galaxy groups at high redshift
is not calibrated in the current framework.

\begin{figure}
\includegraphics[width=8cm]{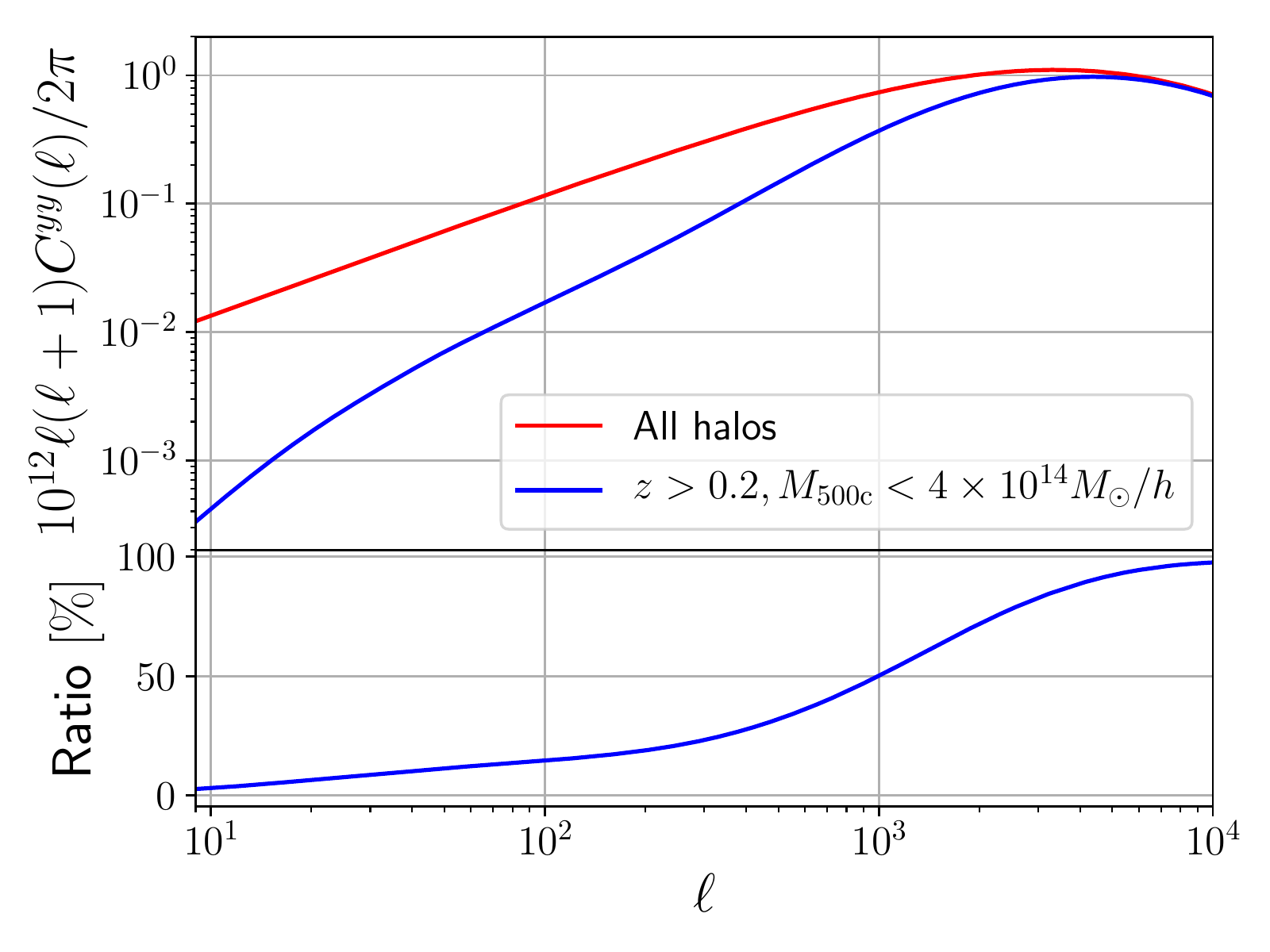}
\includegraphics[width=8cm]{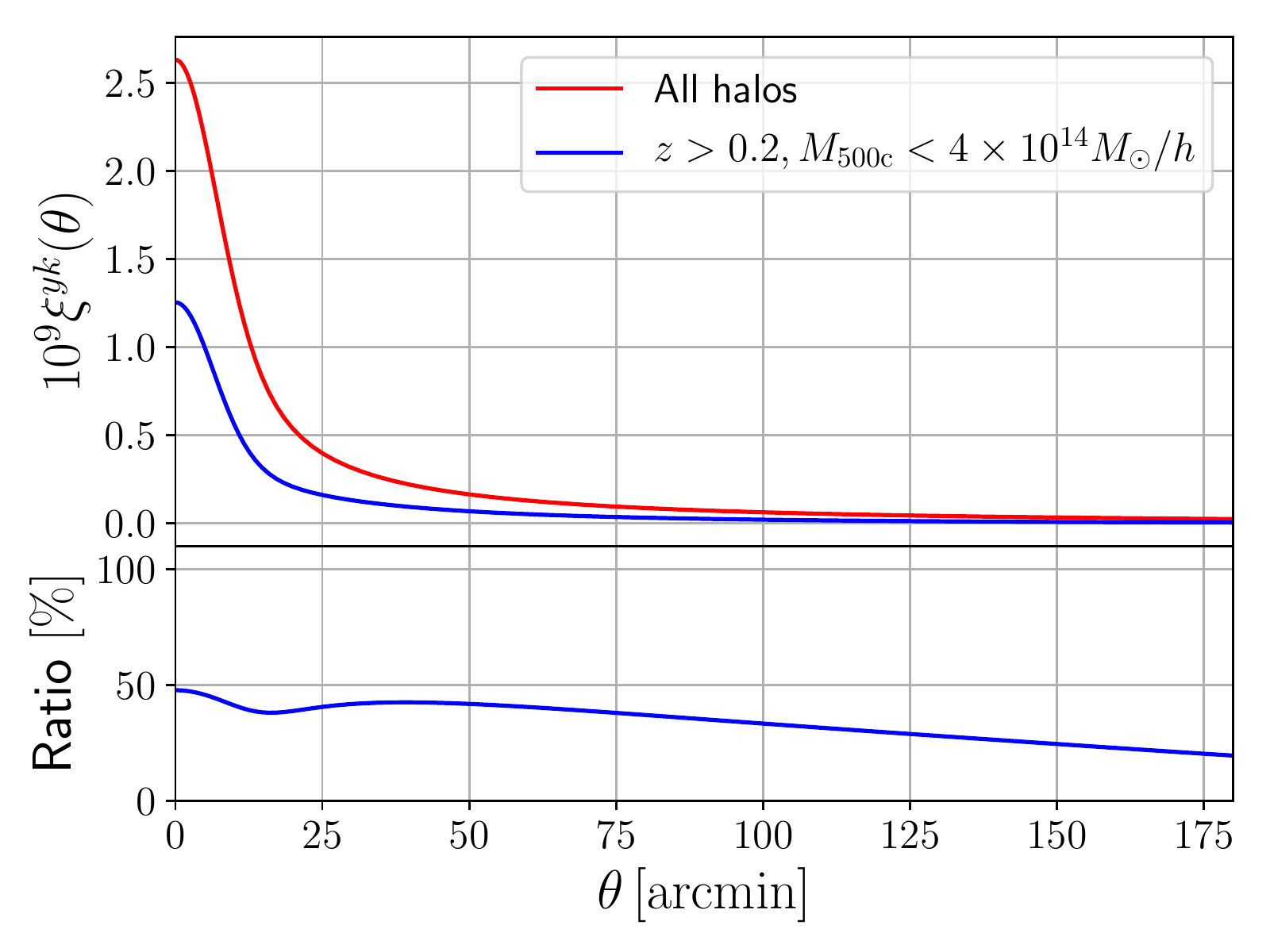}
\caption{Contributions of the power spectrum and cross-correlation
from high redshift groups ($z > 0.2$ and $M_\mathrm{500c} < 4 \times 10^{14} \Msun/h$).
The lower panels show the fraction with respect to contributions from all halos}
\label{fig:highz}
\end{figure}

High-redshift, low-mass groups and clusters contribute a considerable fraction
to the total tSZ power spectrum and tSZ-WL cross-correlation,
as shown in Figure~\ref{fig:highz}, where we show the contribution
from objects with $z > 0.2$ and $M_\mathrm{500c} < 4 \times 10^{14} \Msun/h$.
These objects contribute around $50\%$ to the measured tSZ-WL cross-correlation,
and $50\%$--$100\%$ to the tSZ power spectrum at $1000 < \ell < 10000$.
At multipoles probed by {\it Planck} ($50 \simlt \ell \simlt 1000$),
they contribute still $\sim 10\%$--$50\%$.

\begin{figure}
\includegraphics[width=8cm]{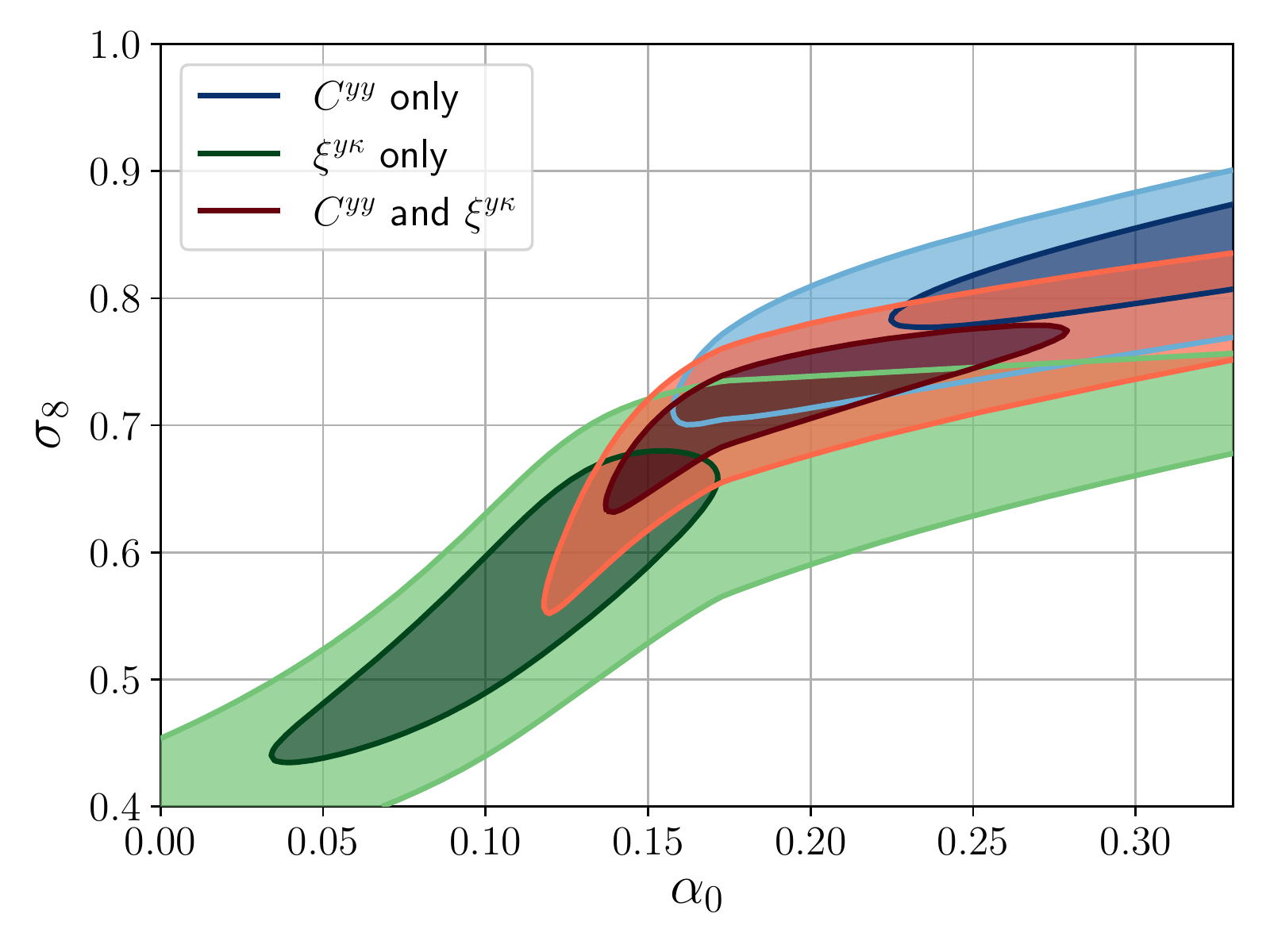}
\caption{Posterior distributions of non-thermal pressure parameters $\alpha_0$
and $\sigma_8$ with $S_* = 0.7$.
The inner (outer) colored region correponds to $1\sigma$ ($2\sigma$)
confidence region.
The results with data sets of power spectra and cross-correlations,
power spectra only, and cross-correlations only are shown in solid red, blue, and green
regions, respectively.}
\label{fig:cr_highS}
\end{figure}

In order to mitigate the tension between data sets,
we consider the case of varying a parameter, $S_*$,
which is the power index of the stellar-to-halo-mass relation
defined in Eq.~\ref{eq:shmr}.
When we take high $S_*$, the gas fraction reduces especially for group size halos,
and then the resultant power spectra and cross-correlations are suppressed.
To demonstrate that the high $S_*$ model has a possibility to alleviate the tension,
we repeat our analysis with $S_* = 0.7$.
In this case, we find that the tension between the two data sets,
the tSZ power spectrum and tSZ-WL cross-correlation,
is mitigated (see Figure~\ref{fig:cr_highS}).
Both data sets are consistent with the fiducial value $\alpha_0 = 0.18$.

We note that the high value for the slope in the stellar fraction, $S_* = 0.7$,
is inconsistent with the results from \citet{Flender2017},
who find $S_* = 0.12 \pm 0.1$. On the other hand,
the steep slope is consistent with the results from \citet{Gonzalez2007},
who analyze the stellar content of groups and clusters
over a wide range of masses, $6 \times 10^{13}$--$10^{15} \Msun$,
and find $S_* = 0.64 \pm 0.13$.

We have also tried other modification to the gas model
in order to mitigate the tension, varying $f_*$, $\epsilon_\mathrm{f}$,
or introducing an additional redshift dependence to the tSZ signal,
but found that enhanced star formation due to high $S_*$ works best,
since it has the most impact on small scales, where the tension originates.

\section{Conclusions}
\label{sec:conclusions}
The tSZ effect probes the thermal properties of the hot, ionized gas in the Universe,
while the WL signal reflects mostly the dark matter distribution,
and is thus less affected by baryonic physics.
Current and future CMB and galaxy redshift surveys enable
measurements of these observables,
which can be used to infer astrophysical and cosmological information.
The first detections of the tSZ-WL cross-correlation have been
recently reported in \citet{VanWaerbeke2014, Hill2014, Hojjati2017}.
The cross-correlation can be a valuable probe in addition to tSZ and WL alone,
as it can help break parameter degeneracies.

In this work, we have modeled the tSZ-WL cross-correlation
using the halo model approach.
We have modeled the pressure profile following
the semi-analytic ICM model from \citet{Flender2017},
as well as the universal pressure profile calibrated
against observations \citep{PlanckCollaboration2013}.
In order to estimate the covariance matrix,
we have produced mock tSZ and WL maps.
For WL, we employ the ray-tracing technique to generate mock maps of the convergence field.
For tSZ, we follow the approach from \citet{Roncarelli2007},
painting the signal into the halos in the simulation.

We constrain the free parameters in our model,
taking into account measurements of the tSZ power spectrum
from {\it Planck} \citep{Planck2015tSZ},
as well as measurements of the tSZ-WL cross-correlation
from RCSLenS and {\it Planck} \citep{Hojjati2017}.
With the observationally calibrated universal pressure profile
from \citet{PlanckCollaboration2013}, and leaving $\sigma_8$
as the only free parameter, we find that the tSZ data alone
prefers $\sigma_8 = 0.785^{+0.029}_{-0.043}$,
consistent with the value of 0.76 found in \citet{Planck2015tSZ}.
However, the value for $\sigma_8$ becomes lower
when taking into account the tSZ-WL cross-correlation.
With the cross-correlation alone we find $\sigma_8 = 0.677^{+0.046}_{-0.077}$,
and with the combined data $\sigma_8 = 0.746^{+0.026}_{-0.038}$.

We repeat the analysis using the pressure profile
from the semi-analytic model from \citet{Flender2017},
leaving the amplitude of non-thermal pressure, $\alpha_0$, and $\sigma_8$
as free parameters.
Here, we find that the tSZ power spectrum prefers $\alpha_0 \sim 0.3$
and $\sigma_8 \sim 0.85$, while the tSZ-WL cross-correlation prefers
a significantly lower $\alpha_0$ of $\sim 0.05$ and $\sigma_8 \sim 0.6$
(see Figure~\ref{fig:cr}).
Ignoring the small scales ($<10\,\mathrm{arcmin}$)
in the analysis seems to alleviate the tension (see Figure~\ref{fig:cr_cut}).

Another way to alleviate the tension between the two data sets is to consider
modifications in the gas model.
We find that allowing for a steep slope
in the stellar-mass-halo-mass relation, $S_* = 0.7$, results in posterior distributions
from the two data sets that are less in tension (see Figure~\ref{fig:cr_highS}),
pointing towards enhanced star formation in low-mass halos.
With the combined data, we find that a model with $\sigma_8 \sim 0.7$, $\alpha_0 \sim 0.2$
is preferred.

The tSZ power spectrum and the tSZ-WL cross-correlation are
exciting new probes of cluster astrophysics and cosmology.
Upcoming galaxy redshift surveys, such as HSC and LSST, and CMB experiments,
such as ACTPol, SPT-3G, and CMB-Stage\,IV, will enable
more precise measurements of these observables,
especially at smaller scales than the $\sim 10\,\mathrm{arcmin}$ size of the {\it Planck} beam,
which currently limits our analysis.
Considering the high quality and wide coverage of future data,
semi-analytic modeling in combination with all-sky simulations \citep{Shirasaki2015}
will be a promising modeling approach.
If future data confirm the tension seen here with higher significance,
we might derive interesting conclusions about the gas physics of groups and clusters,
such as enhanced star formation, i.e. reduced gas content in low-mass halos.
Another possibility would be to derive constraints on the shape of
the pressure profile \citep[see,][]{Battaglia2017}.

\section*{Acknowledgements}
The authors acknowledge Erwin Lau, Nick Battaglia, Hironao Miyatake
and the anonymous referee for useful discussions.
KO and MS are supported by Research Fellowships of the Japan Society for the Promotion of Science
(JSPS) for Young Scientists.
KO was supported by Advanced Leading Graduate Course for Photon Science.
KO, MS and NY acknowledge financial support from JST CREST Grant Number JPMJCR1414.
This work was supported by JSPS Grant-in-Aid for JSPS Research Fellow Grant Number JP16J01512 (KO),
and NSF AST-1412768 (DN).
Argonne National Laboratory's work was supported under the U.S. Department of Energy contract DE-AC02-06CH11357.
Numerical simulations were carried out on Cray XC30 at the Center for Computational Astrophysics,
National Astronomical Observatory of Japan.



\bibliographystyle{mnras}
\bibliography{main}

\begin{thebibliography}{}
\makeatletter
\relax
\def\mn@urlcharsother{\let\do\@makeother \do\$\do\&\do\#\do\^\do\_\do\%\do\~}
\def\mn@doi{\begingroup\mn@urlcharsother \@ifnextchar [ {\mn@doi@}
  {\mn@doi@[]}}
\def\mn@doi@[#1]#2{\def\@tempa{#1}\ifx\@tempa\@empty \href
  {http://dx.doi.org/#2} {doi:#2}\else \href {http://dx.doi.org/#2} {#1}\fi
  \endgroup}
\def\mn@eprint#1#2{\mn@eprint@#1:#2::\@nil}
\def\mn@eprint@arXiv#1{\href {http://arxiv.org/abs/#1} {{\tt arXiv:#1}}}
\def\mn@eprint@dblp#1{\href {http://dblp.uni-trier.de/rec/bibtex/#1.xml}
  {dblp:#1}}
\def\mn@eprint@#1:#2:#3:#4\@nil{\def\@tempa {#1}\def\@tempb {#2}\def\@tempc
  {#3}\ifx \@tempc \@empty \let \@tempc \@tempb \let \@tempb \@tempa \fi \ifx
  \@tempb \@empty \def\@tempb {arXiv}\fi \@ifundefined
  {mn@eprint@\@tempb}{\@tempb:\@tempc}{\expandafter \expandafter \csname
  mn@eprint@\@tempb\endcsname \expandafter{\@tempc}}}

\bibitem[\protect\citeauthoryear{{Aihara} et~al.,}{{Aihara}
  et~al.}{2017}]{Aihara2017}
{Aihara} H.,  et~al., 2017, preprint, \href
  {http://adsabs.harvard.edu/abs/2017arXiv170405858A} {} (\mn@eprint {arXiv}
  {1704.05858})

\bibitem[\protect\citeauthoryear{Arnaud, Pratt, Piffaretti, B{\"{o}}hringer,
  Croston  \& Pointecouteau}{Arnaud et~al.}{2010}]{Arnaud2010}
Arnaud M.,  Pratt G.~W.,  Piffaretti R.,  B{\"{o}}hringer H.,  Croston J.~H.,
  Pointecouteau E.,  2010, \mn@doi [Astronomy and Astrophysics]
  {10.1051/0004-6361/200913416}, 517, A92

\bibitem[\protect\citeauthoryear{Austermann et~al.,}{Austermann
  et~al.}{2012}]{Austermann2012}
Austermann J.~E.,  et~al., 2012, \mn@doi [Millimeter, Submillimeter, and
  Far-Infrared Detectors and Instrumentation for Astronomy VI. Proceedings of
  the SPIE] {10.1117/12.927286}, 8452

\bibitem[\protect\citeauthoryear{Bartelmann \& Schneider}{Bartelmann \&
  Schneider}{2001}]{Bartelmann2001}
Bartelmann M.,  Schneider P.,  2001, \mn@doi [Physics Reports]
  {10.1016/S0370-1573(00)00082-X}, 340, 291

\bibitem[\protect\citeauthoryear{Battaglia, Bond, Pfrommer, Sievers  \&
  Sijacki}{Battaglia et~al.}{2010}]{Battaglia2010}
Battaglia N.,  Bond J.~R.,  Pfrommer C.,  Sievers J.~L.,   Sijacki D.,  2010,
  \mn@doi [The Astrophysical Journal] {10.1088/0004-637X/725/1/91}, 725, 91

\bibitem[\protect\citeauthoryear{Battaglia, Bond, Pfrommer  \&
  Sievers}{Battaglia et~al.}{2012}]{Battaglia2012}
Battaglia N.,  Bond J.~R.,  Pfrommer C.,   Sievers J.~L.,  2012, \mn@doi [The
  Astrophysical Journal] {10.1088/0004-637X/758/2/75}, 758, 75

\bibitem[\protect\citeauthoryear{Battaglia, Hill  \& Murray}{Battaglia
  et~al.}{2015}]{Battaglia2015}
Battaglia N.,  Hill J.~C.,   Murray N.,  2015, \mn@doi [The Astrophysical
  Journal] {10.1088/0004-637X/812/2/154}, 812, 154

\bibitem[\protect\citeauthoryear{{Battaglia}, {Ferraro}, {Schaan}  \&
  {Spergel}}{{Battaglia} et~al.}{2017}]{Battaglia2017}
{Battaglia} N.,  {Ferraro} S.,  {Schaan} E.,   {Spergel} D.,  2017, preprint,
  \href {http://adsabs.harvard.edu/abs/2017arXiv170505881B} {} (\mn@eprint
  {arXiv} {1705.05881})

\bibitem[\protect\citeauthoryear{{Battye}, {Charnock}  \& {Moss}}{{Battye}
  et~al.}{2015}]{Battye2015}
{Battye} R.~A.,  {Charnock} T.,   {Moss} A.,  2015, \mn@doi [\prd]
  {10.1103/PhysRevD.91.103508}, \href
  {http://adsabs.harvard.edu/abs/2015PhRvD..91j3508B} {91, 103508}

\bibitem[\protect\citeauthoryear{Behroozi, Wechsler  \& Wu}{Behroozi
  et~al.}{2013}]{Behroozi2013}
Behroozi P.~S.,  Wechsler R.~H.,   Wu H.-Y.,  2013, \mn@doi [The Astrophysical
  Journal] {10.1088/0004-637X/762/2/109}, 762, 109

\bibitem[\protect\citeauthoryear{Birkinshaw}{Birkinshaw}{1999}]{Birkinshaw1999}
Birkinshaw M.,  1999, \mn@doi [Physics Reports]
  {10.1016/S0370-1573(98)00080-5}, 310, 97

\bibitem[\protect\citeauthoryear{{Bleem} et~al.,}{{Bleem}
  et~al.}{2015}]{Bleem2015}
{Bleem} L.~E.,  et~al., 2015, \mn@doi [The Astrophysical Journal Supplement]
  {10.1088/0067-0049/216/2/27}, \href
  {http://adsabs.harvard.edu/abs/2015ApJS..216...27B} {216, 27}

\bibitem[\protect\citeauthoryear{{Bocquet}, {Saro}, {Dolag}  \&
  {Mohr}}{{Bocquet} et~al.}{2016}]{Bocquet2016}
{Bocquet} S.,  {Saro} A.,  {Dolag} K.,   {Mohr} J.~J.,  2016, \mn@doi [Monthly
  Notices of the Royal Astronomical Society] {10.1093/mnras/stv2657}, \href
  {http://adsabs.harvard.edu/abs/2016MNRAS.456.2361B} {456, 2361}

\bibitem[\protect\citeauthoryear{Bryan \& Norman}{Bryan \&
  Norman}{1998}]{Bryan1998}
Bryan G.~L.,  Norman M.~L.,  1998, \mn@doi [The Astrophysical Journal]
  {10.1086/305262}, 495, 80

\bibitem[\protect\citeauthoryear{Carlstrom, Holder  \& Reese}{Carlstrom
  et~al.}{2002}]{Carlstrom2002}
Carlstrom J.~E.,  Holder G.~P.,   Reese E.~D.,  2002, \mn@doi [Annual Review of
  Astronomy and Astrophysics] {10.1146/annurev.astro.40.060401.093803}, 40, 643

\bibitem[\protect\citeauthoryear{Carlstrom et~al.,}{Carlstrom
  et~al.}{2011}]{Carlstrom2011}
Carlstrom J.~E.,  et~al., 2011, \mn@doi [Publications of the Astronomical
  Society of Pacific] {10.1086/659879}, 123, 568

\bibitem[\protect\citeauthoryear{Cole \& Kaiser}{Cole \&
  Kaiser}{1988}]{Cole1988}
Cole S.,  Kaiser N.,  1988, \mn@doi [Monthly Notices of the Royal Astronomical
  Society] {10.1093/mnras/233.3.637}, 233, 637

\bibitem[\protect\citeauthoryear{{Dark Energy Survey Collaboration}}{{Dark
  Energy Survey Collaboration}}{2016}]{DarkEnergySurveyCollaboration2016}
{Dark Energy Survey Collaboration} 2016, \mn@doi [Monthly Notices of the Royal
  Astronomical Society] {10.1093/mnras/stw641}, 460, 1270

\bibitem[\protect\citeauthoryear{Dolag, Komatsu  \& Sunyaev}{Dolag
  et~al.}{2016}]{Dolag2016}
Dolag K.,  Komatsu E.,   Sunyaev R.,  2016, \mn@doi [Monthly Notices of the
  Royal Astronomical Society] {10.1093/mnras/stw2035}, 463, 1797

\bibitem[\protect\citeauthoryear{Duffy, Schaye, Kay  \& {Dalla Vecchia}}{Duffy
  et~al.}{2008}]{Duffy2008}
Duffy A.~R.,  Schaye J.,  Kay S.~T.,   {Dalla Vecchia} C.,  2008, \mn@doi
  [Monthly Notices of the Royal Astronomical Society Letters]
  {10.1111/j.1745-3933.2008.00537.x}, 390, L64

\bibitem[\protect\citeauthoryear{{Flender}, {Nagai}  \& {McDonald}}{{Flender}
  et~al.}{2017}]{Flender2017}
{Flender} S.,  {Nagai} D.,   {McDonald} M.,  2017, \mn@doi [The Astrophysical
  Journal] {10.3847/1538-4357/aa60bf}, \href
  {http://adsabs.harvard.edu/abs/2017ApJ...837..124F} {837, 124}

\bibitem[\protect\citeauthoryear{George et~al.,}{George
  et~al.}{2015}]{George2015}
George E.~M.,  et~al., 2015, \mn@doi [The Astrophysical Journal]
  {10.1088/0004-637X/799/2/177}, 799, 177

\bibitem[\protect\citeauthoryear{{Gonzalez}, {Zaritsky}  \&
  {Zabludoff}}{{Gonzalez} et~al.}{2007}]{Gonzalez2007}
{Gonzalez} A.~H.,  {Zaritsky} D.,   {Zabludoff} A.~I.,  2007, \mn@doi [The
  Astrophysical Journal] {10.1086/519729}, \href
  {http://adsabs.harvard.edu/abs/2007ApJ...666..147G} {666, 147}

\bibitem[\protect\citeauthoryear{{Hamana} \& {Mellier}}{{Hamana} \&
  {Mellier}}{2001}]{Hamana2001}
{Hamana} T.,  {Mellier} Y.,  2001, \mn@doi [Monthly Notices of the Royal
  Astronomical Society] {10.1046/j.1365-8711.2001.04685.x}, \href
  {http://adsabs.harvard.edu/abs/2001MNRAS.327..169H} {327, 169}

\bibitem[\protect\citeauthoryear{Harnois-D{\'{e}}raps
  et~al.,}{Harnois-D{\'{e}}raps et~al.}{2016}]{Harnois-Deraps2016}
Harnois-D{\'{e}}raps J.,  et~al., 2016, \mn@doi [Monthly Notices of the Royal
  Astronomical Society] {10.1093/mnras/stw947}, 460, 434

\bibitem[\protect\citeauthoryear{{Hasselfield} et~al.,}{{Hasselfield}
  et~al.}{2013}]{Hasselfield2013}
{Hasselfield} M.,  et~al., 2013, \mn@doi [Journal of Cosmology and
  Astroparticle Physics] {10.1088/1475-7516/2013/07/008}, \href
  {http://adsabs.harvard.edu/abs/2013JCAP...07..008H} {7, 008}

\bibitem[\protect\citeauthoryear{Hilbert, Hartlap, White  \& Schneider}{Hilbert
  et~al.}{2009}]{Hilbert2009}
Hilbert S.,  Hartlap J.,  White S. D.~M.,   Schneider P.,  2009, \mn@doi
  [Astronomy and Astrophysics] {10.1051/0004-6361/200811054}, 499, 31

\bibitem[\protect\citeauthoryear{{Hildebrandt} et~al.,}{{Hildebrandt}
  et~al.}{2016}]{Hildebrandt2016}
{Hildebrandt} H.,  et~al., 2016, \mn@doi [Monthly Notices of the Royal
  Astronomical Society] {10.1093/mnras/stw2013}, \href
  {http://adsabs.harvard.edu/abs/2016MNRAS.463..635H} {463, 635}

\bibitem[\protect\citeauthoryear{Hill \& Spergel}{Hill \&
  Spergel}{2014}]{Hill2014}
Hill J.~C.,  Spergel D.~N.,  2014, \mn@doi [Journal of Cosmology and
  Astroparticle Physics] {10.1088/1475-7516/2014/02/030}, 02, 030

\bibitem[\protect\citeauthoryear{Hojjati, McCarthy, Harnois-Deraps, Ma,
  Waerbeke, Hinshaw  \& Brun}{Hojjati et~al.}{2015}]{Hojjati2015}
Hojjati A.,  McCarthy I.~G.,  Harnois-Deraps J.,  Ma Y.-Z.,  Waerbeke L.~V.,
  Hinshaw G.,   Brun A. M.~L.,  2015, \mn@doi [Journal of Cosmology and
  Astroparticle Physics] {10.1088/1475-7516/2015/10/047}, 10, 047

\bibitem[\protect\citeauthoryear{{Hojjati} et~al.,}{{Hojjati}
  et~al.}{2017}]{Hojjati2017}
{Hojjati} A.,  et~al., 2017, \mn@doi [\mnras] {10.1093/mnras/stx1659}, \href
  {http://adsabs.harvard.edu/abs/2017MNRAS.471.1565H} {471, 1565}

\bibitem[\protect\citeauthoryear{{Horowitz} \& {Seljak}}{{Horowitz} \&
  {Seljak}}{2017}]{Horowitz2017}
{Horowitz} B.,  {Seljak} U.,  2017, \mn@doi [Monthly Notices of the Royal
  Astronomical Society] {10.1093/mnras/stx766}, \href
  {http://adsabs.harvard.edu/abs/2017MNRAS.469..394H} {469, 394}

\bibitem[\protect\citeauthoryear{Itoh, Kohyama  \& Nozawa}{Itoh
  et~al.}{1998}]{Itoh1998}
Itoh N.,  Kohyama Y.,   Nozawa S.,  1998, \mn@doi [The Astrophysical Journal]
  {10.1086/305876}, 502, 7

\bibitem[\protect\citeauthoryear{{Khatri} \& {Sunyaev}}{{Khatri} \&
  {Sunyaev}}{2015}]{Khatri2015}
{Khatri} R.,  {Sunyaev} R.,  2015, \mn@doi [\jcap]
  {10.1088/1475-7516/2015/08/013}, \href
  {http://adsabs.harvard.edu/abs/2015JCAP...08..013K} {8, 013}

\bibitem[\protect\citeauthoryear{Kilbinger}{Kilbinger}{2015}]{Kilbinger2015}
Kilbinger M.,  2015, \mn@doi [Reports on Progress in Physics]
  {10.1088/0034-4885/78/8/086901}, 78, 086901

\bibitem[\protect\citeauthoryear{Kitayama}{Kitayama}{2014}]{Kitayama2014}
Kitayama T.,  2014, \mn@doi [Progress of Theoretical and Experimental Physics]
  {10.1093/ptep/ptu055}, 2014, 6B111

\bibitem[\protect\citeauthoryear{{K{\"o}hlinger} et~al.,}{{K{\"o}hlinger}
  et~al.}{2017}]{Kohlinger2017}
{K{\"o}hlinger} F.,  et~al., 2017, \mn@doi [\mnras] {10.1093/mnras/stx1820},
  \href {http://adsabs.harvard.edu/abs/2017MNRAS.471.4412K} {471, 4412}

\bibitem[\protect\citeauthoryear{Komatsu \& Kitayama}{Komatsu \&
  Kitayama}{1999}]{Komatsu1999}
Komatsu E.,  Kitayama T.,  1999, \mn@doi [The Astrophysical Journal]
  {10.1086/312364}, 526, L1

\bibitem[\protect\citeauthoryear{Komatsu \& Seljak}{Komatsu \&
  Seljak}{2001}]{Komatsu2001}
Komatsu E.,  Seljak U.,  2001, \mn@doi [Monthly Notices of the Royal
  Astronomical Society] {10.1046/j.1365-8711.2001.04838.x}, 327, 1353

\bibitem[\protect\citeauthoryear{Komatsu \& Seljak}{Komatsu \&
  Seljak}{2002}]{Komatsu2002}
Komatsu E.,  Seljak U.,  2002, \mn@doi [Monthly Notices of the Royal
  Astronomical Society] {10.1046/j.1365-8711.2002.05889.x}, 336, 1256

\bibitem[\protect\citeauthoryear{{LSST Science Collaboration}}{{LSST Science
  Collaboration}}{2009}]{LSST2009}
{LSST Science Collaboration} 2009, preprint, \href
  {http://adsabs.harvard.edu/abs/2009arXiv0912.0201L} {} (\mn@eprint {arXiv}
  {0912.0201})

\bibitem[\protect\citeauthoryear{{Leauthaud} et~al.,}{{Leauthaud}
  et~al.}{2017}]{Leauthaud2017}
{Leauthaud} A.,  et~al., 2017, \mn@doi [\mnras] {10.1093/mnras/stx258}, \href
  {http://adsabs.harvard.edu/abs/2017MNRAS.467.3024L} {467, 3024}

\bibitem[\protect\citeauthoryear{Ma, {Van Waerbeke}, Hinshaw, Hojjati, Scott
  \& Zuntz}{Ma et~al.}{2015}]{Ma2015}
Ma Y.-Z.,  {Van Waerbeke} L.,  Hinshaw G.,  Hojjati A.,  Scott D.,   Zuntz J.,
  2015, \mn@doi [Journal of Cosmology and Astroparticle Physics]
  {10.1088/1475-7516/2015/09/046}, 9, 046

\bibitem[\protect\citeauthoryear{McCarthy, {Le Brun}, Schaye  \&
  Holder}{McCarthy et~al.}{2014}]{McCarthy2014}
McCarthy I.~G.,  {Le Brun} A. M.~C.,  Schaye J.,   Holder G.~P.,  2014, \mn@doi
  [Monthly Notices of the Royal Astronomical Society] {10.1093/mnras/stu543},
  440, 3645

\bibitem[\protect\citeauthoryear{Munshi, Valageas, van Waerbeke  \&
  Heavens}{Munshi et~al.}{2008}]{Munshi2008}
Munshi D.,  Valageas P.,  van Waerbeke L.,   Heavens A.,  2008, \mn@doi
  [Physics Reports] {10.1016/j.physrep.2008.02.003}, 462, 67

\bibitem[\protect\citeauthoryear{{Munshi}, {Joudaki}, {Coles}, {Smidt}  \&
  {Kay}}{{Munshi} et~al.}{2014}]{Munshi2014}
{Munshi} D.,  {Joudaki} S.,  {Coles} P.,  {Smidt} J.,   {Kay} S.~T.,  2014,
  \mn@doi [\mnras] {10.1093/mnras/stu794}, \href
  {http://adsabs.harvard.edu/abs/2014MNRAS.442...69M} {442, 69}

\bibitem[\protect\citeauthoryear{Nagai, Kravtsov  \& Vikhlinin}{Nagai
  et~al.}{2007}]{Nagai2007}
Nagai D.,  Kravtsov A.~V.,   Vikhlinin A.,  2007, \mn@doi [The Astrophysical
  Journal] {10.1086/521328}, 668, 1

\bibitem[\protect\citeauthoryear{Navarro, Frenk  \& White}{Navarro
  et~al.}{1996}]{Navarro1996}
Navarro J.~F.,  Frenk C.~S.,   White S. D.~M.,  1996, \mn@doi [The
  Astrophysical Journal] {10.1086/177173}, 462, 563

\bibitem[\protect\citeauthoryear{Navarro, Frenk  \& White}{Navarro
  et~al.}{1997}]{Navarro1997}
Navarro J.~F.,  Frenk C.~S.,   White S. D.~M.,  1997, \mn@doi [The
  Astrophysical Journal] {10.1086/304888}, 490, 493

\bibitem[\protect\citeauthoryear{Nelson, Lau  \& Nagai}{Nelson
  et~al.}{2014}]{Nelson2014}
Nelson K.,  Lau E.~T.,   Nagai D.,  2014, \mn@doi [The Astrophysical Journal]
  {10.1088/0004-637X/792/1/25}, 792, 25

\bibitem[\protect\citeauthoryear{Niemack et~al.,}{Niemack
  et~al.}{2010}]{Niemack2010}
Niemack M.~D.,  et~al., 2010, \mn@doi [Proceedings of the SPIE]
  {10.1117/12.857464}, 7741

\bibitem[\protect\citeauthoryear{Nishimichi et~al.,}{Nishimichi
  et~al.}{2009}]{Nishimichi2009}
Nishimichi T.,  et~al., 2009, \mn@doi [Publications of the Astronomical Society
  of Japan] {10.1093/pasj/61.2.321}, 61, 321

\bibitem[\protect\citeauthoryear{Nishimichi, Taruya, Koyama  \&
  Sabiu}{Nishimichi et~al.}{2010}]{Nishimichi2010}
Nishimichi T.,  Taruya A.,  Koyama K.,   Sabiu C.,  2010, \mn@doi [Journal of
  Cosmology and Astroparticle Physics] {10.1088/1475-7516/2010/07/002}, 07, 002

\bibitem[\protect\citeauthoryear{Nozawa, Itoh  \& Kohyama}{Nozawa
  et~al.}{1998}]{Nozawa1998}
Nozawa S.,  Itoh N.,   Kohyama Y.,  1998, \mn@doi [The Astrophysical Journal]
  {10.1086/306401}, 508, 17

\bibitem[\protect\citeauthoryear{Oguri \& Takada}{Oguri \&
  Takada}{2011}]{Oguri2011}
Oguri M.,  Takada M.,  2011, \mn@doi [Physical Review D]
  {10.1103/PhysRevD.83.023008}, 83, 023008

\bibitem[\protect\citeauthoryear{Ostriker, Bode  \& Babul}{Ostriker
  et~al.}{2005}]{Ostriker2005}
Ostriker J.~P.,  Bode P.,   Babul A.,  2005, \mn@doi [The Astrophysical
  Journal] {10.1086/497122}, 634, 964

\bibitem[\protect\citeauthoryear{{Planck Collaboration}}{{Planck
  Collaboration}}{2013}]{PlanckCollaboration2013}
{Planck Collaboration} 2013, \mn@doi [Astronomy and Astrophysics]
  {10.1051/0004-6361/201220040}, 550, A131

\bibitem[\protect\citeauthoryear{{Planck Collaboration}}{{Planck
  Collaboration}}{2016a}]{Planck2015parameters}
{Planck Collaboration} 2016a, \mn@doi [Astronomy and Astrophysics]
  {10.1051/0004-6361/201525830}, 594, A13

\bibitem[\protect\citeauthoryear{{Planck Collaboration}}{{Planck
  Collaboration}}{2016b}]{Planck2015tSZ}
{Planck Collaboration} 2016b, \mn@doi [Astronomy and Astrophysics]
  {10.1051/0004-6361/201525826}, 594, A22

\bibitem[\protect\citeauthoryear{Roncarelli, Moscardini, Borgani  \&
  Dolag}{Roncarelli et~al.}{2007}]{Roncarelli2007}
Roncarelli M.,  Moscardini L.,  Borgani S.,   Dolag K.,  2007, \mn@doi [Monthly
  Notices of the Royal Astronomical Society]
  {10.1111/j.1365-2966.2007.11914.x}, 378, 1259

\bibitem[\protect\citeauthoryear{Sato, Hamana, Takahashi, Takada, Yoshida,
  Matsubara  \& Sugiyama}{Sato et~al.}{2009}]{Sato2009}
Sato M.,  Hamana T.,  Takahashi R.,  Takada M.,  Yoshida N.,  Matsubara T.,
  Sugiyama N.,  2009, \mn@doi [The Astrophysical Journal]
  {10.1088/0004-637X/701/2/945}, 701, 945

\bibitem[\protect\citeauthoryear{Sehgal, Bode, Das, Hernandez-Monteagudo,
  Huffenberger, Lin, Ostriker  \& Trac}{Sehgal et~al.}{2010}]{Sehgal2010}
Sehgal N.,  Bode P.,  Das S.,  Hernandez-Monteagudo C.,  Huffenberger K.,  Lin
  Y.-T.,  Ostriker J.~P.,   Trac H.,  2010, \mn@doi [The Astrophysical Journal]
  {10.1088/0004-637X/709/2/920}, 709, 920

\bibitem[\protect\citeauthoryear{Shaw, Nagai, Bhattacharya  \& Lau}{Shaw
  et~al.}{2010}]{Shaw2010}
Shaw L.~D.,  Nagai D.,  Bhattacharya S.,   Lau E.~T.,  2010, \mn@doi [The
  Astrophysical Journal] {10.1088/0004-637X/725/2/1452}, 725, 1452

\bibitem[\protect\citeauthoryear{Shirasaki, Hamana  \& Yoshida}{Shirasaki
  et~al.}{2015}]{Shirasaki2015}
Shirasaki M.,  Hamana T.,   Yoshida N.,  2015, \mn@doi [Monthly Notices of the
  Royal Astronomical Society] {10.1093/mnras/stv1854}, 453, 3043

\bibitem[\protect\citeauthoryear{Sievers et~al.,}{Sievers
  et~al.}{2013}]{Sievers2013}
Sievers J.~L.,  et~al., 2013, \mn@doi [Journal of Cosmology and Astroparticle
  Physics] {10.1088/1475-7516/2013/10/060}, 10, 060

\bibitem[\protect\citeauthoryear{Springel}{Springel}{2005}]{Springel2005}
Springel V.,  2005, \mn@doi [Monthly Notices of the Royal Astronomical Society]
  {10.1111/j.1365-2966.2005.09655.x}, 364, 1105

\bibitem[\protect\citeauthoryear{{Sunyaev} \& {Zeldovich}}{{Sunyaev} \&
  {Zeldovich}}{1972}]{Sunyaev1972}
{Sunyaev} R.~A.,  {Zeldovich} Y.~B.,  1972, Comments on Astrophysics and Space
  Physics, \href {http://adsabs.harvard.edu/abs/1972CoASP...4..173S} {4, 173}

\bibitem[\protect\citeauthoryear{{Sunyaev} \& {Zeldovich}}{{Sunyaev} \&
  {Zeldovich}}{1980}]{Sunyaev1980}
{Sunyaev} R.~A.,  {Zeldovich} Y.~B.,  1980, \mn@doi [Monthly Notices of the
  Royal Astronomical Society] {10.1093/mnras/190.3.413}, \href
  {http://adsabs.harvard.edu/abs/1980MNRAS.190..413S} {190, 413}

\bibitem[\protect\citeauthoryear{Swetz et~al.,}{Swetz et~al.}{2011}]{Swetz2011}
Swetz D.~S.,  et~al., 2011, \mn@doi [The Astrophysical Journal Supplement]
  {10.1088/0067-0049/194/2/41}, 194, 41

\bibitem[\protect\citeauthoryear{Tinker, Robertson, Kravtsov, Klypin, Warren,
  Yepes  \& Gottl{\"{o}}ber}{Tinker et~al.}{2010}]{Tinker2010}
Tinker J.~L.,  Robertson B.~E.,  Kravtsov A.~V.,  Klypin A.,  Warren M.~S.,
  Yepes G.,   Gottl{\"{o}}ber S.,  2010, \mn@doi [The Astrophysical Journal]
  {10.1088/0004-637X/724/2/878}, 724, 878

\bibitem[\protect\citeauthoryear{Trac, Bode  \& Ostriker}{Trac
  et~al.}{2011}]{Trac2011}
Trac H.,  Bode P.,   Ostriker J.~P.,  2011, \mn@doi [The Astrophysical Journal]
  {10.1088/0004-637X/727/2/94}, 727, 94

\bibitem[\protect\citeauthoryear{Tr{\"{o}}ster \& van Waerbeke}{Tr{\"{o}}ster
  \& van Waerbeke}{2014}]{Troster2014}
Tr{\"{o}}ster T.,  van Waerbeke L.,  2014, \mn@doi [Journal of Cosmology and
  Astroparticle Physics] {10.1088/1475-7516/2014/11/008}, 11, 008

\bibitem[\protect\citeauthoryear{Ursino, Galeazzi  \& Roncarelli}{Ursino
  et~al.}{2010}]{Ursino2010}
Ursino E.,  Galeazzi M.,   Roncarelli M.,  2010, \mn@doi [The Astrophysical
  Journal] {10.1088/0004-637X/721/1/46}, 721, 46

\bibitem[\protect\citeauthoryear{Valageas \& Nishimichi}{Valageas \&
  Nishimichi}{2011}]{Valageas2011}
Valageas P.,  Nishimichi T.,  2011, \mn@doi [Astronomy and Astrophysics]
  {10.1051/0004-6361/201015685}, 527, A87

\bibitem[\protect\citeauthoryear{White \& Hu}{White \& Hu}{2000}]{White2000}
White M.,  Hu W.,  2000, \mn@doi [The Astrophysical Journal] {10.1086/309009},
  537, 1

\bibitem[\protect\citeauthoryear{{van Waerbeke}}{{van
  Waerbeke}}{2000}]{VanWaerbeke2000}
{van Waerbeke} L.,  2000, \mn@doi [Monthly Notices of the Royal Astronomical
  Society] {10.1046/j.1365-8711.2000.03259.x}, \href
  {http://adsabs.harvard.edu/abs/2000MNRAS.313..524V} {313, 524}

\bibitem[\protect\citeauthoryear{{van Waerbeke}, Hinshaw  \& Murray}{{van
  Waerbeke} et~al.}{2014}]{VanWaerbeke2014}
{van Waerbeke} L.,  Hinshaw G.,   Murray N.,  2014, \mn@doi [Physical Review D]
  {10.1103/PhysRevD.89.023508}, 89, 023508

\makeatother
\end{thebibliography}



\appendix
\section{Summary of symbols}
\label{sec:symbols}
In Table~\ref{tab:symbols}, we summarize symbols used
in this paper.

\begin{table*}
\caption{Symbols used in this paper.}
\label{tab:symbols}
\begin{tabular}{ccc}
\hline
Symbol & Definition & Reference equation \\
\hline
 & Halo model & \\
\hline
$y$ & Compton-$y$ & \eqref{eq:y_def} \\
$P_\mathrm{e}$ & Free electron pressure & \eqref{eq:y_def} \\
$C^{yy} (\ell)$ & Power spectrum of Compton-$y$ & \eqref{eq:yy_hm1} \\
$C^{yy \mathrm{(1h)}} (\ell)$ & 1-halo term of $C^{yy} (\ell)$ & \eqref{eq:yy_hm2} \\
$C^{yy \mathrm{(2h)}} (\ell)$ & 2-halo term of $C^{yy} (\ell)$ & \eqref{eq:yy_hm3} \\
$y_\ell$ & Fourier transform of Compton-$y$ of single halo & \eqref{eq:yy_hm3} \\
$dn/dM$ & Halo mass function & \eqref{eq:yy_hm2}, \eqref{eq:yy_hm3} \\
$b$ & Halo bias & \eqref{eq:yy_hm3} \\
$P_\mathrm{m} (k, z)$ & Linear matter power spectrum & \eqref{eq:yy_hm3} \\
$M_\Delta$ & Halo mass with the overdensity $\Delta$ & \eqref{eq:M_delta} \\
$M_{500\mathrm{c}}$ & Halo mass with the overdensity $\Delta = 500$ & ... \\
$M_\vir$ & Virial halo mass & \eqref{eq:Delta_vir}, \eqref{eq:M_vir} \\
$\rho (r)$ & Density profile of halo & \eqref{eq:rho} \\
$c (M_\vir, z)$ & Concentration parameter & \eqref{eq:conc} \\
$\kappa$ & Weak lensing convergence & ... \\
$\kappa_\ell$ & Fourier transform of convergence of single halo & \eqref{eq:kappa_ell} \\
$C^{y\kappa} (\ell)$ & Cross power spectrum of Compton-$y$ and convergence & \eqref{eq:yk_hm1} \\
$C^{y\kappa \mathrm{(1h)}} (\ell)$ & 1-halo term of $C^{y\kappa} (\ell)$ & \eqref{eq:yk_hm2} \\
$C^{y\kappa \mathrm{(2h)}} (\ell)$ & 2-halo term of $C^{y\kappa} (\ell)$ & \eqref{eq:yk_hm3} \\
$\xi^{y\kappa} (\theta)$ & Cross correlation function of Compton-$y$ and convergence & \eqref{eq:xi_yk} \\
\hline
 & Semi-analytic model of the ICM & \\
\hline
$P_\mathrm{tot} (r)$ & Total pressure profile of halo & \eqref{eq:Euler_eq}, \eqref{eq:P_tot} \\
$P_\mathrm{nt} (r)$ & Non-thermal pressure profile of halo & \eqref{eq:nt_frac} \\
$\rho_g (r)$ & Gas density profile of halo & (\ref{eq:Euler_eq}), \eqref{eq:rho_g} \\
$\alpha (z)$ & Amplitude of radial profile of non-thermal fraction & \eqref{eq:nt_frac}, \eqref{eq:alpha_z} \\
$n_\mathrm{nt}$ & Power law index of non-thermal fraction & \eqref{eq:nt_frac} \\
$\alpha_0$ & Amplitude of $\alpha (z)$ & \eqref{eq:alpha_z} \\
$\beta$ & Parameter which determines redshift dependence of $\alpha(z)$ & \eqref{eq:alpha_z} \\
$\epsilon_\mathrm{DM}$ & Parameter which describes energy transfer between dark matter and gas & ... \\
$\epsilon_\mathrm{f}$ & Parameter which regulates stellar feedback energy & ... \\
$f_*$ & Amplitude of stellar mass fraction relation & \eqref{eq:shmr} \\
$S_*$ & Power law index of stellar mass fraction relation & \eqref{eq:shmr} \\
\hline
\end{tabular}
\end{table*}


\bsp 
\label{lastpage}
\end{document}